\documentclass[
reprint,
superscriptaddress,
 amsmath,amssymb,
 aps,longbibliography,
prl,nofootinbib,
floatfix,
toc=flat,
]{revtex4-1}

\usepackage{graphicx}
\usepackage{placeins}
\usepackage{txfonts}
\usepackage{braket}
\usepackage{xcolor}
\usepackage{nicefrac}

\definecolor{myciteColor}{rgb}{0.0,0.5,0.23}
\usepackage[colorlinks=true,citecolor=myciteColor,linkcolor=myciteColor,urlcolor=myciteColor]{hyperref}

\begin{document}

\title{Observation of a finite-energy phase transition in a one-dimensional quantum simulator}
\author{Alexander Schuckert$^{*\dagger}$}
\affiliation{Joint Quantum Institute and Joint Center for Quantum Information and Computer Science, University of Maryland and NIST, College Park, Maryland 20742, USA}
\author{Or Katz$^{*\dagger}$}
\affiliation{Duke Quantum Center, Department of Physics and Electrical and Computer Engineering, Duke University, Durham, NC 27701}
\author{Lei Feng}
\affiliation{Duke Quantum Center, Department of Physics and Electrical and Computer Engineering, Duke University, Durham, NC 27701}
\author{Eleanor Crane}
\affiliation{Joint Quantum Institute and Joint Center for Quantum Information and Computer Science, University of Maryland and NIST, College Park, Maryland 20742, USA}
\author{Arinjoy De}
\affiliation{Joint Quantum Institute and Joint Center for Quantum Information and Computer Science, University of Maryland and NIST, College Park, Maryland 20742, USA}
\author{Mohammad Hafezi}
\affiliation{Joint Quantum Institute and Joint Center for Quantum Information and Computer Science, University of Maryland and NIST, College Park, Maryland 20742, USA}
\author{Alexey V.~Gorshkov}
\affiliation{Joint Quantum Institute and Joint Center for Quantum Information and Computer Science, University of Maryland and NIST, College Park, Maryland 20742, USA}
\author{Christopher Monroe}
\affiliation{Duke Quantum Center, Department of Physics and Electrical and Computer Engineering, Duke University, Durham, NC 27701}

\def\thefootnote{*}\footnotetext{These authors contributed equally to this work.}
\def\thefootnote{$\dagger$}\footnotetext{Corresponding authors: aschu@umd.edu, or.katz@duke.edu}

\begin{abstract} 
One of the most striking many-body phenomena in nature is the sudden change of macroscopic properties as the temperature or energy reaches a critical value. Such equilibrium transitions have been predicted and observed in two and three spatial dimensions, but have long been thought not to exist in one-dimensional (1D) systems. Fifty years ago, Dyson and Thouless pointed out that a phase transition in 1D can occur in the presence of long-range interactions, but an experimental realization has so far not been achieved due to the requirement to both prepare equilibrium states and realize sufficiently long-range interactions. Here we report on the first experimental demonstration of a finite-energy phase transition in 1D.  
We use the simple observation that finite-energy states can be prepared by time-evolving product initial states and letting them thermalize under the dynamics of a many-body Hamiltonian.
By preparing initial states with different energies in a 1D trapped-ion quantum simulator, we study the finite-energy phase diagram of a long-range interacting quantum system. We observe a ferromagnetic equilibrium phase transition as well as a crossover from a low-energy polarized paramagnet to a high-energy unpolarized paramagnet in a system of up to $23$ spins, in excellent agreement with numerical simulations.
Our work demonstrates the ability of quantum simulators to realize and study previously inaccessible phases at finite energy density. 
\end{abstract} 

\maketitle

Equilibrium phase transitions underlie many quantum phenomena in nature, from the creation of primordial fluctuations in the early universe~\cite{Kibble1976,Zurek1985} to the melting of confined hadrons into the quark-gluon plasma~\cite{Fukushima2010} and the emergence of a superconducting state at high temperatures in the cuprates~\cite{Keimer2015}. Equilibrium phase transitions require the presence of both ordered and disordered phases, which have been observed in two and three spatial dimensions. In one-dimensional (1D) systems, disordered phases often have a lower free energy than ordered ones, leading to the absence of phase transitions. This is because the entropy gained by destroying the order is larger than the energy cost~\cite{vanHove1950, Landau1980}. More than half a century ago, Dyson and Thouless argued that this energy cost can outweigh the entropy gain if the interactions are sufficiently long-ranged~\cite{Dyson1969,Thouless1969}. However, despite the extensive theoretical work on phase transitions in 1D long-range systems since these seminal works~\cite{RevModPhys.95.035002}, an experimental realization of this prediction has so far not been achieved. Recently, the advent of quantum simulators has enabled the study of highly controlled long-range interacting systems. This has lead to the discovery of many exotic quantum phenomena, including non-local spreading of quasiparticles~\cite{Jurcevic2014}, dynamical quantum phase transitions~\cite{Jurcevic2017, Zhang20172}, time crystals~\cite{Zhang2017,Choi2017}, continuous symmetry breaking~\cite{Lei2022,Chen2023}, supersolidity~\cite{PhysRevLett.122.130405, PhysRevX.9.011051}, and superdiffusive spin transport~\cite{Joshi2022}. Of the experimental platforms used in those studies, only trapped ions have interactions that are in principle long-range enough to observe an equilibrium phase transition in 1D. However, preparing equilibrium states in spin-system quantum simulators such as trapped ions and even digital quantum computers has been challenging~\cite{Poulin2009,Temme2011,Chowdhury2017,Motta2019,Zhu2020,Lu2021,shtanko2021,tazhigulov2022,Schuckert2022,hemery2023}.

    Here, we use the intrinsic many-body equilibration within sub-systems of isolated quantum systems to prepare equilibrium states in a long-range interacting quantum system realized by trapped ions. Our method relies on the foundational understanding of quantum thermalization offered by the eigenstate thermalization hypothesis (ETH)~\cite{Srednicki1996,Rigol2008}, whose validity and limitations have been studied both numerically~\cite{Rigol2008} and experimentally~\cite{Trotzky2012, Kaufman2016}. We first prepare non-equilibrium states at a range of different energies, and let these states thermalize under the dynamics of a 1D many-body Hamiltonian with programmable long-range interactions. We then use these thermalized states to measure the order parameters of ferromagnetic and paramagnetic phases as a function of energy density. This enables us to study the existence of a possible equilibrium phase transition in a 1D spin system.

\begin{figure*}
     \centering
     \includegraphics[width=6in]{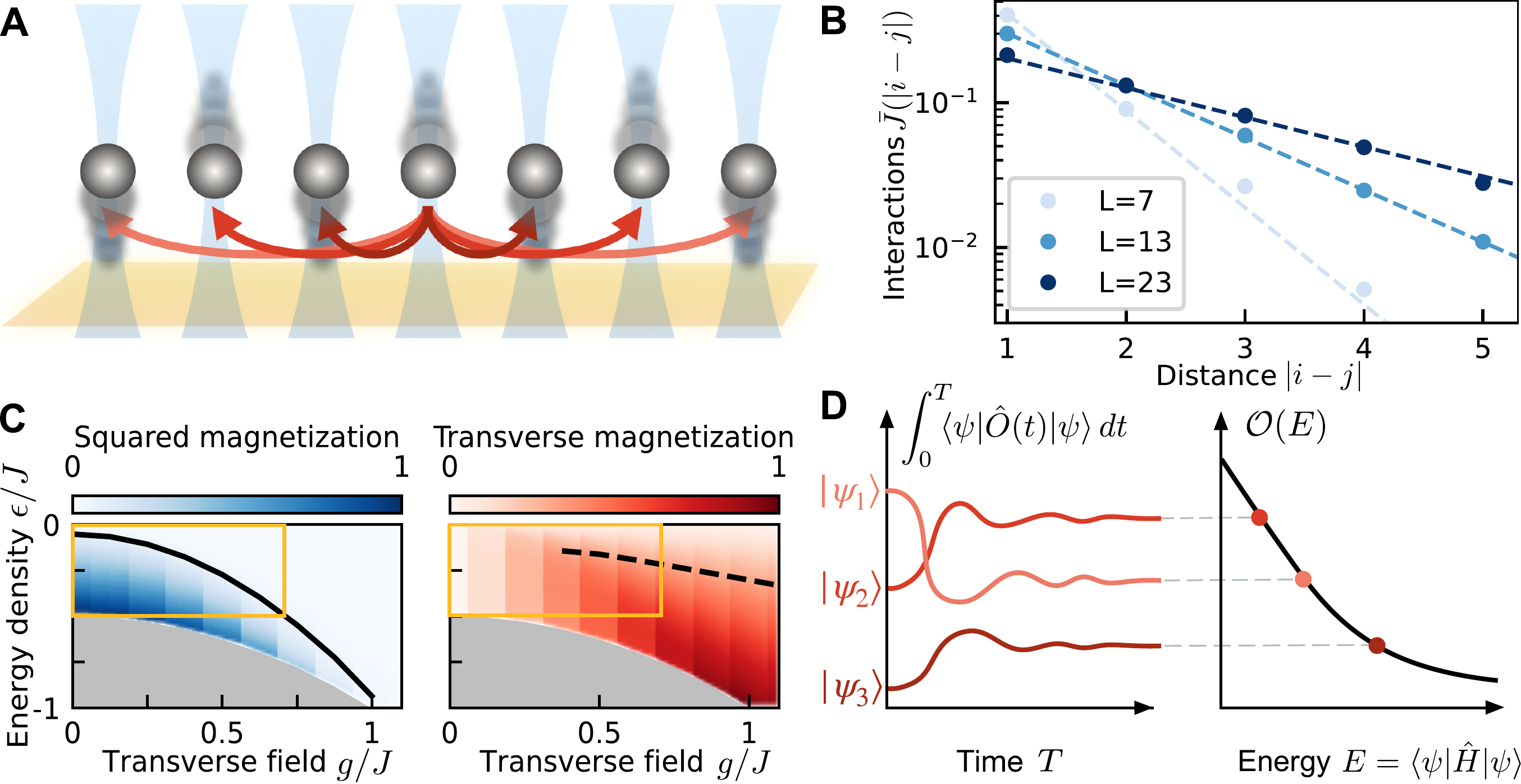}
     \caption{\textbf{Observing an equilibrium phase transition in a one-dimensional chain of ions.} (\textbf{A}) Ions are confined in a chain using a chip trap (yellow). Individual Raman laser beams (blue) couple an internal spin-$1/2$ degree of freedom of the ions to the lowest-energy motional mode of the ion crystal (gray), mediating exponentially decaying interactions between the spins (red). (\textbf{B}) Calculated experimental interaction strengths (colored dots), averaged over the center-of-mass coordinate $\bar J(l)=\frac{1}{ L-l}\sum_i J_{i,i+l}$,  along with the target interactions in Eq.~(\ref{eq:Hamiltonian}), see SM~\cite{supp}, section I for how we calculate the $J_{ij}$.  (\textbf{C}) Equilibrium phase boundary of the model in Eq.~\eqref{eq:Hamiltonian} (solid line), as well as the squared magnetization (blue shading) and transverse magnetization (red shading, dashed line indicates where transverse magnetization is equal to $0.75$). We extrapolated to the infinite-system-size limit from finite-size matrix-product-state simulations. Gray area indicates energies with no states. The gold frame is the regime experimentally probed in this work.
     (\textbf{D}) The time-averaged expectation value of a time-evolved local observable at late times gives an estimate for the equilibrium expectation value at the energy of the initial state. Repeating this experiment for initial states with different energy results in an estimate for the value of the observable in equilibrium as a function of energy.}
     \label{fig:1}
 \end{figure*}

\section{Realizing a long-range interacting many-body system}

In our experiment, we encode a pseudo-spin $1/2$ in the electronic ground-state levels $\ket{\uparrow}=\ket{F = 1, M = 0}$ and $\ket{\downarrow}=\ket{F = 0, M = 0}$ of ${}^{171}$Yb${}^+$ ions confined in a linear Paul trap on a chip~\cite{katz2023}, as illustrated in Fig.~\ref{fig:1}A. We apply a tighly focussed individual addressing beam on each ion and a globally-addressing wide beam on all ions to drive the transition between the spin levels via a stimulated Raman process. These transitions couple off-resonantly to the phonon modes of the ion chain, driving simultaneously and nearly symmetrically the red- and blue-sideband transitions, resulting in effective Ising interactions between the spins. We realize interactions that decay exponentially with ion separation distance by choosing a Raman beat-note frequency which has a detuning from the carrier frequency that is close to the lowest radial phonon-mode frequency. We then remove the alternating sign of the interactions by spatially staggering the phases of the Raman beams, see supplementary materials (SM)~\cite{supp}, section I. Control over the optical frequencies of the individual beams sets an effective magnetic field in the transverse direction~\cite{Porras2004, Monroe2022}. In total, this approximately realizes the many-body spin Hamiltonian
\begin{equation}
    \hat H = -\frac{J}{2\mathcal{N}} \sum_{i<j} \exp\left(-\tilde \gamma|i-j|\right)\hat \sigma^x_i \hat \sigma^x_j - g \sum_i \hat \sigma^z_i,
    \label{eq:Hamiltonian}
\end{equation}
where $\hat \sigma^\alpha_i$ are the Pauli matrices, $g$ is the transverse-field strength, and $i, j$ run over integers from $1$ to $L$ for an ion chain of length $L$. In order to enable the possibility of a phase transition in a system with exponentially decaying interactions,  we decrease the decay rate of the interactions with increasing system size by choosing $\tilde \gamma=\gamma/L$ with $\gamma=10.8$, see Fig.~\ref{fig:1}B. This effectively gives our system the characteristics of a long-range interacting model. To render the energy extensive~\cite{Kac1963}, we rescale the overall prefactor of the Hamiltonian by  $\mathcal{N}=\frac{1}{L-1}\sum_{i< j}\exp\left(-\tilde \gamma|i-j|\right)$.

 \begin{figure*}
     \centering
     \includegraphics[width=6in]{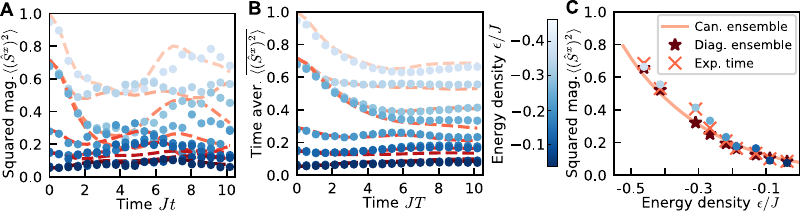}
     \caption{\textbf{Verification of equilibration.} (\textbf{A}) Time-evolved squared magnetization in the experiment (dots) and numerical simulations (dashed lines). (\textbf{B}) Time-average (up to time $T$) of the time-evolved squared magnetization using the data from (\textbf{A}) (dots) and the corresponding numerical data (dashed lines), evaluated according to Eq.~\eqref{eq:timeaver}. The color bar indicates the energy density of the initial states.  (\textbf{C}) Comparison of the latest-time experimental data points from (\textbf{B}) (dots), numerical data evolved until the experimental time (crosses) and to infinite time, i.e. the diagonal ensemble (stars, evaluated according to the RHS in Eq.~\eqref{eq:timeaver}). The expectation from the canonical ensemble is shown as a solid line. The numerics use the experimentally realized interactions, see SM~\cite{supp}, section I. $L=13$, $g=0.31J$. Error bars for the experimental data are quantum projection noise and are smaller than the point size.}
     \label{fig:2}
 \end{figure*}

To confirm that the Hamiltonian in Eq.~\eqref{eq:Hamiltonian} exhibits an equilibrium phase transition, we use matrix product state simulations, see SM~\cite{supp}, section II. Indeed, for $g\leq J$, we find a low-energy ferromagnetic phase, characterized by a non-zero squared magnetization $\braket{\hat S_x^2}\equiv\sum_{ij}\braket{\hat \sigma^x_i\hat\sigma^x_j}/L^2$ (Fig.~\ref{fig:1}C). The paramagnet outside this ordered phase consists of two regions that are connected by a crossover: at low energies, the system is polarized along the transverse field, while at high energies both the transverse magnetization $\braket{\hat S_z}\equiv\sum_{i}\braket{\hat \sigma^z_i}/L$ and the squared magnetization vanish, and the system is effectively in an unpolarized mixed state. While we expect the universality class to be given by the all-to-all connected model $\gamma=0$, we numerically found that the critical temperature depends on $\gamma$, see SM~\cite{supp}, section II.

\section{Probing finite-energy states in a quantum simulator of spins}

Our goal is to study the equilibrium phase diagram in Fig.~\ref{fig:1}C in a trapped-ion simulator. However, the preparation of equilibrium states by thermalization with an external bath is challenging in spin-system simulators such as trapped ions. This is due to the existence of  noise sources whose effect can be modelled as a coupling to an infinite temperature bath, which leads to a trivial non-equilibrium steady state. Instead, we use the fact that subsystems of a many-body quantum system thermalize under the system's own dynamics due to the eigenstate thermalization hypothesis~\cite{Srednicki1996}: the expectation values of local observables $\braket{n|\hat O|n}$  with respect to eigenstates $\ket{n}$ of chaotic many-body Hamiltonians coincide with their value in the microcanonical ensemble $\mathcal{O}(E_n)$, evaluated at the corresponding eigenenergy $E_n$.  When starting from an initial state $\ket{\psi}$ with an average energy $E=\braket{\psi|\hat H|\psi}$, the time-averaged observable until time $T$, 
\begin{equation}
    \overline{\braket{\psi| \hat O(t) |\psi} }\equiv \int^T_0 dt \braket{\psi| \hat O(t) |\psi} \stackrel{T\rightarrow \infty} {\rightarrow}\sum_{n} |\braket{n|\psi}|^2 \braket{n|\hat O|n}, \label{eq:timeaver}
\end{equation}
therefore coincides with the microcanonical ensemble $\mathcal{O}(E)$ if the energy-density variance of the initial state $(\braket{\psi|\hat H^2|\psi}-\braket{\psi|\hat H|\psi}^2)/L^2$ vanishes as $L\rightarrow \infty$, i.e.~if it fulfills the condition for a proper thermodynamic ensemble. This condition is fulfilled for most physical initial states~\cite{Rigol2008} and, in these cases, $|\braket{n|\psi}|^2$ is called the diagonal ensemble. The eigenstate thermalization hypothesis therefore motivates the following simple prescription to evaluating equilibrium observables (see Fig.~\ref{fig:1}D): we prepare initial states with different energies $E$ and evolve them to sufficiently late times while measuring the observable $\hat O$. Finally, we record the time-averaged late-time observables on the left-hand side of Eq.~\eqref{eq:timeaver} as a function of $E$ as the resulting estimate for $\mathcal{O}(E)$.

\begin{figure*}[ht!]
     \centering
     \includegraphics[width=4.75in]{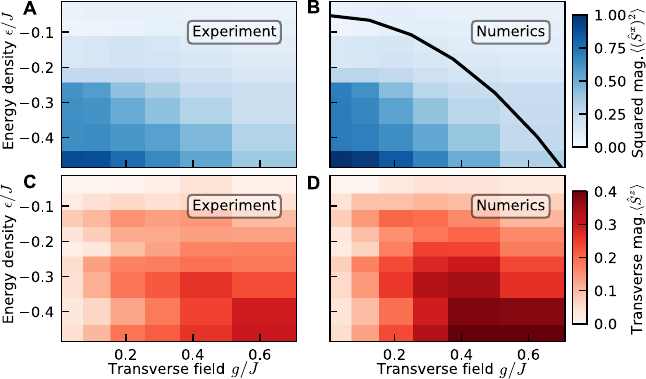}
     \caption{\textbf{Equilibrium phase diagram.} The probed region corresponds to the gold frame in Fig.~\ref{fig:1}C. (\textbf{A} and \textbf{B}) Squared magnetization and (\textbf{C} and \textbf{D}) transverse magnetization from (\textbf{A} and \textbf{C}) the experiment   and (\textbf{B} and \textbf{D}) numerics. Numerics are obtained by evolving to the same time as the experiment. $L=13$. The transverse fields are $g/J=0.04, 0.10, 0.21, 0.31, 0.41,
       0.62$. The black line is the phase transition line from Fig.~\ref{fig:1}C. }
     \label{fig:3}
 \end{figure*}

The energy range that can be probed in this scheme depends on the initial states $\ket{\psi}$. We use product states in the $\hat \sigma^x$ basis. They are the eigenstates of the Hamiltonian for vanishing transverse field $g=0$. The lowest-energy state is the maximally polarized state, and the energy density is mainly controlled by the number of spin flips. We refer the reader to the SM~\cite{supp}, section I, for the specific states we choose. These product states cover a large range of energies even for non-vanishing $g/J$, indicated by the horizontal edges of the gold frame in Fig.~\ref{fig:1}C. The energy density variance of these product states is given by $g^2/L$, therefore also fulfilling the requirement on $|\braket{n|\psi}|^2$ imposed by the eigenstate thermalization hypothesis. 

To test the practical operation of our scheme, we measure the squared magnetization as a function of time for several product initial states for a chain of $L=13$ spins (Fig.~\ref{fig:2}A, dots). We find excellent agreement with the exact numerical solution (dashed lines), which does not include experimental imperfections except the inhomogeneity of the interactions. Evaluating the time average in Eq.~\eqref{eq:timeaver}, we find an approximate convergence with the averaging time $T$ for $JT\gtrsim 8$. Finally, we show the latest-time values of the time-averaged squared magnetization in Fig.~\ref{fig:2}C, along with the numerical results from time evolving to the same time as the experiment (crosses) and to infinite time (diagonal ensemble, stars). The canonical ensemble is shown as a solid line. 
We find good  agreement between the equilibrium diagonal and canonical ensembles and the experimental data, see SM~\cite{supp}, section III, for a more in-depth comparison. This confirms that this scheme enables the evaluation of equilibrium observables in our model on timescales accessible to the experiment.

\begin{figure*}[ht!]
     \centering
     \includegraphics[width=5.5in]{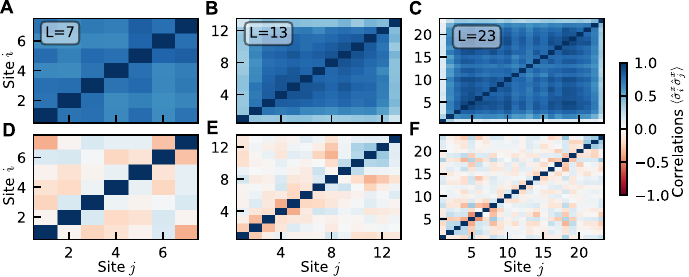}
     \caption{\textbf{Correlations.}  Experimentally measured correlations at late times (not time averaged) at (\textbf{A}-\textbf{C}) low energy ($\epsilon/J\approx-0.43,-0.46,-0.48$) and (\textbf{D}-\textbf{F}) high energy ($\epsilon/J\approx 0, -0.04, 0.04$) for $L=7$, $g/J=0.24$ (left column), $L=13$, $g/J=0.31$ (middle column), $L=23$, $g/J=0.18$ (right column). Times are $Jt\approx 6.3, 10.2, 8.2$ for $L=7,13,23$. For the corresponding numerical result, see SM~\cite{supp}, section II.}
     \label{fig:4}
 \end{figure*}

\section{Observing a finite-energy phase transition}

Having validated our scheme to prepare equilibrium states, we now use it as a tool to probe the phase diagram in Fig.~\ref{fig:1}C by repeating the procedure for many values of the magnetic field $g/J$. We display the result for the squared magnetization in Fig.~\ref{fig:3}A along with the numerical result, which was obtained by time evolving the initial states to the same time as the experiment. We find good agreement between the two. At low energy densities and small transverse fields, a large squared magnetization is observed. Conversely, at high energy densities and large transverse fields, the squared magnetization is small. This is consistent with a phase transition between a ferromagnet and an unmagnetized state. In particular,  we find a good qualitative match between our finite-size experiment and the infinite-system-size extrapolation displayed in the gold frame in Fig.~\ref{fig:1}C, including in particular the phase transition line (black line in  Fig.~\ref{fig:3}B). This indicates weak finite-size effects. More specifically, we also measured the squared magnetization for varying system sizes, indeed finding a qualitative match between them, with some residual finite-size dependence due to the slightly differing transverse field $g/J$ and the inhomogeneity of the experimentally realized interactions, see SM~\cite{supp}, section III.

To probe this transition further, we also measure the transverse magnetization, shown in Fig.~\ref{fig:3}C and D, again finding reasonable agreement between numerics and experiment. Importantly, we find a large transverse magnetization for large transverse fields and low energies, indicating a polarized state along the transverse field. This polarization is destroyed as we move to higher energy densities, which is indicative of a finite-energy crossover from a polarized paramagnet in the ground state at large transverse fields to an unpolarized state at high energies. The presence of this crossover results from the fact that, for large transverse field $g/J$, the energy density is proportional to the transverse magnetization such that  $\braket{\hat S^z}\rightarrow 0$ for $(\epsilon/J)\rightarrow 0$. In other words, the single-site reduced density matrix crosses over from a pure product state at low energies to an effective completely mixed state at high energies.

Counterintuitively, for intermediate energy scales around $\epsilon/J\approx -0.3$, the transverse magnetization displays a maximum as the field is increased, and then decreases again as the field is increased further. This is due to the fact that our initial states have energy densities $\epsilon/J\lesssim 1/2$ (gold frame in Fig.~\ref{fig:1}C) while the ground state energy decreases with $g$. This means that, as the transverse field is increased, we probe higher energy densities relative to the ground state energy (c.f. the gold frame in Fig.\ref{fig:1}C), explaining the decrease of the transverse magnetization.

Ordered phases are accompanied by long-range correlations throughout the system. We measured $\braket{\hat \sigma^x_i\hat \sigma^x_j}$ for three different system sizes at the lowest and highest energy densities, see Fig.~\ref{fig:4}. We again find good agreement with numerical simulations, see SM~\cite{supp}, section II. We find large positive correlations in the bulk of the system for low energies, confirming the presence of a ferromagnetically ordered state. At high energies, the correlations approximately vanish throughout the system, showing that the high-energy state is disordered. 

\section{Discussion and Outlook}

In this work, we have for the first time observed a finite-energy phase transition in one spatial dimension. To do so, we prepared equilibrium states in a trapped-ion quantum simulator using a scheme based on the intrinsic thermalization of closed quantum many-body systems. This required time evolution to relatively late times, and our experiment agreed well with numerical calculations. For large systems, this method would be challenging to simulate numerically. Similarly, trotterized digital time evolution would be difficult in this long-range model due to the necessity to apply $L^2$ entangling gates per Trotter step.

In the future, this scheme can be immediately applied in a variety of analog and digital spin-simulator platforms including Rydberg atoms, quantum dots, and ultracold polar molecules to probe equilibrium states. For instance, equilibrium phases with continuous symmetry breaking~\cite{Chen2023,Lei2022,Yang2023} and other phases of various spin models \cite{gong2016kaleidoscope,defenu2023long} could be studied. While we focused on finite-energy measurements, the effective temperature of the late-time state can in principle be extracted using fluctuation-dissipation relations~\cite{Hartke2020, Schuckert2020} to enable direct comparison with solid-state experiments at finite temperature. This way, long-standing questions in condensed matter physics such as the nature of the excitations above  spin-liquid ground states~\cite{Semeghini2021,Satzinger2021} could be studied in quantum simulators. 

\section{Acknowledgments}
 We acknowledge discussions with Henrik Dreyer, Khaldoon Ghanem, Kévin Hémery, Mikhail D. Lukin and Torsten V. Zache.
 
 \textbf{Funding.} This work is supported by the NSF STAQ Program (PHY-1818914), the DOE Quantum Systems Accelerator (DE-FOA-0002253), the AFOSR MURI on Dissipative Quantum Control (FA9550-19-1-0399). A.S., E.C., and A.V.G.~were also supported in part by the NSF QLCI (award No.~OMA-2120757), the DoE ASCR Quantum Testbed Pathfinder program (awards No.~DE-SC0019040 and No.~DE-SC0024220), DoE ASCR Accelerated Research in Quantum Computing program (award No.~DE-SC0020312), AFOSR, ARO MURI, AFOSR MURI, and the DARPA SAVaNT ADVENT program.     
 
 \textbf{Author contributions.} A.S.\ and O.K.\ devised the research. O.K., L.F., A.D., C.M.\ contributed to the experimental setup. O.K., L.F.\ performed experiments. A.S.\ performed  numerical simulations. A.S., O.K., L.F.\ analyzed the experimental data. A.S., E.C., M.H., A.G.\ contributed to the theoretical analysis. A.G.\ and C.M.\ supervised the research. A.S.\ wrote the initial manuscript, and all authors contributed revisions.

\textbf{Competing interests.}  C.M. is a founder of IonQ, Inc.\ and has a personal financial interest in the company. All other authors declare no competing interests.

\textbf{Data and materials availability.} All data and codes are available from the corresponding authors upon reasonable request.
 \FloatBarrier
 \bibliography{lit.bib}

\begin{thebibliography}{45}%
\makeatletter
\providecommand \@ifxundefined [1]{%
 \@ifx{#1\undefined}
}%
\providecommand \@ifnum [1]{%
 \ifnum #1\expandafter \@firstoftwo
 \else \expandafter \@secondoftwo
 \fi
}%
\providecommand \@ifx [1]{%
 \ifx #1\expandafter \@firstoftwo
 \else \expandafter \@secondoftwo
 \fi
}%
\providecommand \natexlab [1]{#1}%
\providecommand \enquote  [1]{``#1''}%
\providecommand \bibnamefont  [1]{#1}%
\providecommand \bibfnamefont [1]{#1}%
\providecommand \citenamefont [1]{#1}%
\providecommand \href@noop [0]{\@secondoftwo}%
\providecommand \href [0]{\begingroup \@sanitize@url \@href}%
\providecommand \@href[1]{\@@startlink{#1}\@@href}%
\providecommand \@@href[1]{\endgroup#1\@@endlink}%
\providecommand \@sanitize@url [0]{\catcode `\\12\catcode `\$12\catcode
  `\&12\catcode `\#12\catcode `\^12\catcode `\_12\catcode `\%12\relax}%
\providecommand \@@startlink[1]{}%
\providecommand \@@endlink[0]{}%
\providecommand \url  [0]{\begingroup\@sanitize@url \@url }%
\providecommand \@url [1]{\endgroup\@href {#1}{\urlprefix }}%
\providecommand \urlprefix  [0]{URL }%
\providecommand \Eprint [0]{\href }%
\providecommand \doibase [0]{http://dx.doi.org/}%
\providecommand \selectlanguage [0]{\@gobble}%
\providecommand \bibinfo  [0]{\@secondoftwo}%
\providecommand \bibfield  [0]{\@secondoftwo}%
\providecommand \translation [1]{[#1]}%
\providecommand \BibitemOpen [0]{}%
\providecommand \bibitemStop [0]{}%
\providecommand \bibitemNoStop [0]{.\EOS\space}%
\providecommand \EOS [0]{\spacefactor3000\relax}%
\providecommand \BibitemShut  [1]{\csname bibitem#1\endcsname}%
\let\auto@bib@innerbib\@empty
\bibitem [{\citenamefont {Kibble}(1976)}]{Kibble1976}%
  \BibitemOpen
  \bibfield  {author} {\bibinfo {author} {\bibfnamefont {T~W~B}\ \bibnamefont
  {Kibble}},\ }\bibfield  {title} {\enquote {\bibinfo {title} {Topology of
  cosmic domains and strings},}\ }\href {\doibase 10.1088/0305-4470/9/8/029}
  {\bibfield  {journal} {\bibinfo  {journal} {Journal of Physics A:
  Mathematical and General}\ }\textbf {\bibinfo {volume} {9}},\ \bibinfo
  {pages} {1387--1398} (\bibinfo {year} {1976})}\BibitemShut {NoStop}%
\bibitem [{\citenamefont {Zurek}(1985)}]{Zurek1985}%
  \BibitemOpen
  \bibfield  {author} {\bibinfo {author} {\bibfnamefont {W.~H.}\ \bibnamefont
  {Zurek}},\ }\bibfield  {title} {\enquote {\bibinfo {title} {Cosmological
  experiments in superfluid helium?}}\ }\href {\doibase 10.1038/317505a0}
  {\bibfield  {journal} {\bibinfo  {journal} {Nature}\ }\textbf {\bibinfo
  {volume} {317}},\ \bibinfo {pages} {505--508} (\bibinfo {year}
  {1985})}\BibitemShut {NoStop}%
\bibitem [{\citenamefont {Fukushima}\ and\ \citenamefont
  {Hatsuda}(2010)}]{Fukushima2010}%
  \BibitemOpen
  \bibfield  {author} {\bibinfo {author} {\bibfnamefont {Kenji}\ \bibnamefont
  {Fukushima}}\ and\ \bibinfo {author} {\bibfnamefont {Tetsuo}\ \bibnamefont
  {Hatsuda}},\ }\bibfield  {title} {\enquote {\bibinfo {title} {The phase
  diagram of dense {QCD}},}\ }\href {\doibase 10.1088/0034-4885/74/1/014001}
  {\bibfield  {journal} {\bibinfo  {journal} {Reports on Progress in Physics}\
  }\textbf {\bibinfo {volume} {74}},\ \bibinfo {pages} {014001} (\bibinfo
  {year} {2010})}\BibitemShut {NoStop}%
\bibitem [{\citenamefont {Keimer}\ \emph {et~al.}(2015)\citenamefont {Keimer},
  \citenamefont {Kivelson}, \citenamefont {Norman}, \citenamefont {Uchida},\
  and\ \citenamefont {Zaanen}}]{Keimer2015}%
  \BibitemOpen
  \bibfield  {author} {\bibinfo {author} {\bibfnamefont {B.}~\bibnamefont
  {Keimer}}, \bibinfo {author} {\bibfnamefont {S.~A.}\ \bibnamefont
  {Kivelson}}, \bibinfo {author} {\bibfnamefont {M.~R.}\ \bibnamefont
  {Norman}}, \bibinfo {author} {\bibfnamefont {S.}~\bibnamefont {Uchida}}, \
  and\ \bibinfo {author} {\bibfnamefont {J.}~\bibnamefont {Zaanen}},\
  }\bibfield  {title} {\enquote {\bibinfo {title} {From quantum matter to
  high-temperature superconductivity in copper oxides},}\ }\href {\doibase
  10.1038/nature14165} {\bibfield  {journal} {\bibinfo  {journal} {Nature}\
  }\textbf {\bibinfo {volume} {518}},\ \bibinfo {pages} {179--186} (\bibinfo
  {year} {2015})}\BibitemShut {NoStop}%
\bibitem [{\citenamefont {van Hove}(1950)}]{vanHove1950}%
  \BibitemOpen
  \bibfield  {author} {\bibinfo {author} {\bibfnamefont {L.}~\bibnamefont {van
  Hove}},\ }\bibfield  {title} {\enquote {\bibinfo {title} {{Sur
  L{\textquotesingle}int{\'{e}}grale de Configuration Pour Les Syst{\`{e}}mes
  De Particules {\`{A}} Une Dimension}},}\ }\href {\doibase
  10.1016/0031-8914(50)90072-3} {\bibfield  {journal} {\bibinfo  {journal}
  {Physica}\ }\textbf {\bibinfo {volume} {16}},\ \bibinfo {pages} {137--143}
  (\bibinfo {year} {1950})}\BibitemShut {NoStop}%
\bibitem [{\citenamefont {Landau}\ and\ \citenamefont
  {Lifshitz}(1980)}]{Landau1980}%
  \BibitemOpen
  \bibfield  {author} {\bibinfo {author} {\bibfnamefont {L.D.}\ \bibnamefont
  {Landau}}\ and\ \bibinfo {author} {\bibfnamefont {E.M.}\ \bibnamefont
  {Lifshitz}},\ }\href {\doibase 10.1016/c2009-0-24487-4} {\emph {\bibinfo
  {title} {{Statistical Physics}}}}\ (\bibinfo  {publisher}
  {Butterworth-Heinemann},\ \bibinfo {year} {1980})\BibitemShut {NoStop}%
\bibitem [{\citenamefont {Dyson}(1969)}]{Dyson1969}%
  \BibitemOpen
  \bibfield  {author} {\bibinfo {author} {\bibfnamefont {Freeman~J.}\
  \bibnamefont {Dyson}},\ }\bibfield  {title} {\enquote {\bibinfo {title}
  {{Existence of a Phase-Transition in a One-Dimensional {{Ising}}
  Ferromagnet}},}\ }\href {\doibase 10.1007/BF01645907} {\bibfield  {journal}
  {\bibinfo  {journal} {Communications in Mathematical Physics}\ }\textbf
  {\bibinfo {volume} {12}},\ \bibinfo {pages} {91--107} (\bibinfo {year}
  {1969})}\BibitemShut {NoStop}%
\bibitem [{\citenamefont {Thouless}(1969)}]{Thouless1969}%
  \BibitemOpen
  \bibfield  {author} {\bibinfo {author} {\bibfnamefont {D.~J.}\ \bibnamefont
  {Thouless}},\ }\bibfield  {title} {\enquote {\bibinfo {title} {{Long-Range
  Order in One-Dimensional Ising Systems}},}\ }\href {\doibase
  10.1103/physrev.187.732} {\bibfield  {journal} {\bibinfo  {journal} {Physical
  Review}\ }\textbf {\bibinfo {volume} {187}},\ \bibinfo {pages} {732--733}
  (\bibinfo {year} {1969})}\BibitemShut {NoStop}%
\bibitem [{\citenamefont {Defenu}\ \emph
  {et~al.}(2023{\natexlab{a}})\citenamefont {Defenu}, \citenamefont {Donner},
  \citenamefont {Macr\`{\i}}, \citenamefont {Pagano}, \citenamefont {Ruffo},\
  and\ \citenamefont {Trombettoni}}]{RevModPhys.95.035002}%
  \BibitemOpen
  \bibfield  {author} {\bibinfo {author} {\bibfnamefont {Nicol\`o}\
  \bibnamefont {Defenu}}, \bibinfo {author} {\bibfnamefont {Tobias}\
  \bibnamefont {Donner}}, \bibinfo {author} {\bibfnamefont {Tommaso}\
  \bibnamefont {Macr\`{\i}}}, \bibinfo {author} {\bibfnamefont {Guido}\
  \bibnamefont {Pagano}}, \bibinfo {author} {\bibfnamefont {Stefano}\
  \bibnamefont {Ruffo}}, \ and\ \bibinfo {author} {\bibfnamefont {Andrea}\
  \bibnamefont {Trombettoni}},\ }\bibfield  {title} {\enquote {\bibinfo {title}
  {Long-range interacting quantum systems},}\ }\href {\doibase
  10.1103/RevModPhys.95.035002} {\bibfield  {journal} {\bibinfo  {journal}
  {Rev. Mod. Phys.}\ }\textbf {\bibinfo {volume} {95}},\ \bibinfo {pages}
  {035002} (\bibinfo {year} {2023}{\natexlab{a}})}\BibitemShut {NoStop}%
\bibitem [{\citenamefont {Jurcevic}\ \emph {et~al.}(2014)\citenamefont
  {Jurcevic}, \citenamefont {Lanyon}, \citenamefont {Hauke}, \citenamefont
  {Hempel}, \citenamefont {Zoller}, \citenamefont {Blatt},\ and\ \citenamefont
  {Roos}}]{Jurcevic2014}%
  \BibitemOpen
  \bibfield  {author} {\bibinfo {author} {\bibfnamefont {P.}~\bibnamefont
  {Jurcevic}}, \bibinfo {author} {\bibfnamefont {B.~P.}\ \bibnamefont
  {Lanyon}}, \bibinfo {author} {\bibfnamefont {P.}~\bibnamefont {Hauke}},
  \bibinfo {author} {\bibfnamefont {C.}~\bibnamefont {Hempel}}, \bibinfo
  {author} {\bibfnamefont {P.}~\bibnamefont {Zoller}}, \bibinfo {author}
  {\bibfnamefont {R.}~\bibnamefont {Blatt}}, \ and\ \bibinfo {author}
  {\bibfnamefont {C.~F.}\ \bibnamefont {Roos}},\ }\bibfield  {title} {\enquote
  {\bibinfo {title} {{Quasiparticle engineering and entanglement propagation in
  a quantum many-body system}},}\ }\href {\doibase 10.1038/nature13461}
  {\bibfield  {journal} {\bibinfo  {journal} {Nature}\ }\textbf {\bibinfo
  {volume} {511}},\ \bibinfo {pages} {202--205} (\bibinfo {year}
  {2014})}\BibitemShut {NoStop}%
\bibitem [{\citenamefont {Jurcevic}\ \emph {et~al.}(2017)\citenamefont
  {Jurcevic}, \citenamefont {Shen}, \citenamefont {Hauke}, \citenamefont
  {Maier}, \citenamefont {Brydges}, \citenamefont {Hempel}, \citenamefont
  {Lanyon}, \citenamefont {Heyl}, \citenamefont {Blatt},\ and\ \citenamefont
  {Roos}}]{Jurcevic2017}%
  \BibitemOpen
  \bibfield  {author} {\bibinfo {author} {\bibfnamefont {P.}~\bibnamefont
  {Jurcevic}}, \bibinfo {author} {\bibfnamefont {H.}~\bibnamefont {Shen}},
  \bibinfo {author} {\bibfnamefont {P.}~\bibnamefont {Hauke}}, \bibinfo
  {author} {\bibfnamefont {C.}~\bibnamefont {Maier}}, \bibinfo {author}
  {\bibfnamefont {T.}~\bibnamefont {Brydges}}, \bibinfo {author} {\bibfnamefont
  {C.}~\bibnamefont {Hempel}}, \bibinfo {author} {\bibfnamefont {B.~P.}\
  \bibnamefont {Lanyon}}, \bibinfo {author} {\bibfnamefont {M.}~\bibnamefont
  {Heyl}}, \bibinfo {author} {\bibfnamefont {R.}~\bibnamefont {Blatt}}, \ and\
  \bibinfo {author} {\bibfnamefont {C.~F.}\ \bibnamefont {Roos}},\ }\bibfield
  {title} {\enquote {\bibinfo {title} {{Direct Observation of Dynamical Quantum
  Phase Transitions in an Interacting Many-Body System}},}\ }\href {\doibase
  10.1103/PhysRevLett.119.080501} {\bibfield  {journal} {\bibinfo  {journal}
  {Phys. Rev. Lett.}\ }\textbf {\bibinfo {volume} {119}},\ \bibinfo {pages}
  {080501} (\bibinfo {year} {2017})}\BibitemShut {NoStop}%
\bibitem [{\citenamefont {Zhang}\ \emph
  {et~al.}(2017{\natexlab{a}})\citenamefont {Zhang}, \citenamefont {Pagano},
  \citenamefont {Hess}, \citenamefont {Kyprianidis}, \citenamefont {Becker},
  \citenamefont {Kaplan}, \citenamefont {Gorshkov}, \citenamefont {Gong},\ and\
  \citenamefont {Monroe}}]{Zhang20172}%
  \BibitemOpen
  \bibfield  {author} {\bibinfo {author} {\bibfnamefont {J.}~\bibnamefont
  {Zhang}}, \bibinfo {author} {\bibfnamefont {G.}~\bibnamefont {Pagano}},
  \bibinfo {author} {\bibfnamefont {P.~W.}\ \bibnamefont {Hess}}, \bibinfo
  {author} {\bibfnamefont {A.}~\bibnamefont {Kyprianidis}}, \bibinfo {author}
  {\bibfnamefont {P.}~\bibnamefont {Becker}}, \bibinfo {author} {\bibfnamefont
  {H.}~\bibnamefont {Kaplan}}, \bibinfo {author} {\bibfnamefont {A.~V.}\
  \bibnamefont {Gorshkov}}, \bibinfo {author} {\bibfnamefont {Z.-X.}\
  \bibnamefont {Gong}}, \ and\ \bibinfo {author} {\bibfnamefont
  {C.}~\bibnamefont {Monroe}},\ }\bibfield  {title} {\enquote {\bibinfo {title}
  {{Observation of a many-body dynamical phase transition with a 53-qubit
  quantum simulator}},}\ }\href {\doibase 10.1038/nature24654} {\bibfield
  {journal} {\bibinfo  {journal} {Nature}\ }\textbf {\bibinfo {volume} {551}},\
  \bibinfo {pages} {601--604} (\bibinfo {year}
  {2017}{\natexlab{a}})}\BibitemShut {NoStop}%
\bibitem [{\citenamefont {Zhang}\ \emph
  {et~al.}(2017{\natexlab{b}})\citenamefont {Zhang}, \citenamefont {Hess},
  \citenamefont {Kyprianidis}, \citenamefont {Becker}, \citenamefont {Lee},
  \citenamefont {Smith}, \citenamefont {Pagano}, \citenamefont {Potirniche},
  \citenamefont {Potter}, \citenamefont {Vishwanath}, \citenamefont {Yao},\
  and\ \citenamefont {Monroe}}]{Zhang2017}%
  \BibitemOpen
  \bibfield  {author} {\bibinfo {author} {\bibfnamefont {J.}~\bibnamefont
  {Zhang}}, \bibinfo {author} {\bibfnamefont {P.~W.}\ \bibnamefont {Hess}},
  \bibinfo {author} {\bibfnamefont {A.}~\bibnamefont {Kyprianidis}}, \bibinfo
  {author} {\bibfnamefont {P.}~\bibnamefont {Becker}}, \bibinfo {author}
  {\bibfnamefont {A.}~\bibnamefont {Lee}}, \bibinfo {author} {\bibfnamefont
  {J.}~\bibnamefont {Smith}}, \bibinfo {author} {\bibfnamefont
  {G.}~\bibnamefont {Pagano}}, \bibinfo {author} {\bibfnamefont {I.-D.}\
  \bibnamefont {Potirniche}}, \bibinfo {author} {\bibfnamefont {A.~C.}\
  \bibnamefont {Potter}}, \bibinfo {author} {\bibfnamefont {A.}~\bibnamefont
  {Vishwanath}}, \bibinfo {author} {\bibfnamefont {N.~Y.}\ \bibnamefont {Yao}},
  \ and\ \bibinfo {author} {\bibfnamefont {C.}~\bibnamefont {Monroe}},\
  }\bibfield  {title} {\enquote {\bibinfo {title} {Observation of a discrete
  time crystal},}\ }\href {\doibase 10.1038/nature21413} {\bibfield  {journal}
  {\bibinfo  {journal} {Nature}\ }\textbf {\bibinfo {volume} {543}},\ \bibinfo
  {pages} {217--220} (\bibinfo {year} {2017}{\natexlab{b}})}\BibitemShut
  {NoStop}%
\bibitem [{\citenamefont {Choi}\ \emph {et~al.}(2017)\citenamefont {Choi},
  \citenamefont {Choi}, \citenamefont {Landig}, \citenamefont {Kucsko},
  \citenamefont {Zhou}, \citenamefont {Isoya}, \citenamefont {Jelezko},
  \citenamefont {Onoda}, \citenamefont {Sumiya}, \citenamefont {Khemani},
  \citenamefont {von Keyserlingk}, \citenamefont {Yao}, \citenamefont
  {Demler},\ and\ \citenamefont {Lukin}}]{Choi2017}%
  \BibitemOpen
  \bibfield  {author} {\bibinfo {author} {\bibfnamefont {Soonwon}\ \bibnamefont
  {Choi}}, \bibinfo {author} {\bibfnamefont {Joonhee}\ \bibnamefont {Choi}},
  \bibinfo {author} {\bibfnamefont {Renate}\ \bibnamefont {Landig}}, \bibinfo
  {author} {\bibfnamefont {Georg}\ \bibnamefont {Kucsko}}, \bibinfo {author}
  {\bibfnamefont {Hengyun}\ \bibnamefont {Zhou}}, \bibinfo {author}
  {\bibfnamefont {Junichi}\ \bibnamefont {Isoya}}, \bibinfo {author}
  {\bibfnamefont {Fedor}\ \bibnamefont {Jelezko}}, \bibinfo {author}
  {\bibfnamefont {Shinobu}\ \bibnamefont {Onoda}}, \bibinfo {author}
  {\bibfnamefont {Hitoshi}\ \bibnamefont {Sumiya}}, \bibinfo {author}
  {\bibfnamefont {Vedika}\ \bibnamefont {Khemani}}, \bibinfo {author}
  {\bibfnamefont {Curt}\ \bibnamefont {von Keyserlingk}}, \bibinfo {author}
  {\bibfnamefont {Norman~Y.}\ \bibnamefont {Yao}}, \bibinfo {author}
  {\bibfnamefont {Eugene}\ \bibnamefont {Demler}}, \ and\ \bibinfo {author}
  {\bibfnamefont {Mikhail~D.}\ \bibnamefont {Lukin}},\ }\bibfield  {title}
  {\enquote {\bibinfo {title} {{Observation of discrete time-crystalline order
  in a disordered dipolar many-body system}},}\ }\href {\doibase
  10.1038/nature21426} {\bibfield  {journal} {\bibinfo  {journal} {Nature}\
  }\textbf {\bibinfo {volume} {543}},\ \bibinfo {pages} {221--225} (\bibinfo
  {year} {2017})}\BibitemShut {NoStop}%
\bibitem [{\citenamefont {Feng}\ \emph {et~al.}(2022)\citenamefont {Feng},
  \citenamefont {Katz}, \citenamefont {Haack}, \citenamefont {Maghrebi},
  \citenamefont {Gorshkov}, \citenamefont {Gong}, \citenamefont {Cetina},\ and\
  \citenamefont {Monroe}}]{Lei2022}%
  \BibitemOpen
  \bibfield  {author} {\bibinfo {author} {\bibfnamefont {Lei}\ \bibnamefont
  {Feng}}, \bibinfo {author} {\bibfnamefont {Or}~\bibnamefont {Katz}}, \bibinfo
  {author} {\bibfnamefont {Casey}\ \bibnamefont {Haack}}, \bibinfo {author}
  {\bibfnamefont {Mohammad}\ \bibnamefont {Maghrebi}}, \bibinfo {author}
  {\bibfnamefont {Alexey~V.}\ \bibnamefont {Gorshkov}}, \bibinfo {author}
  {\bibfnamefont {Zhexuan}\ \bibnamefont {Gong}}, \bibinfo {author}
  {\bibfnamefont {Marko}\ \bibnamefont {Cetina}}, \ and\ \bibinfo {author}
  {\bibfnamefont {Christopher}\ \bibnamefont {Monroe}},\ }\bibfield  {title}
  {\enquote {\bibinfo {title} {{Continuous Symmetry Breaking in a Trapped-Ion
  Spin Chain}},}\ }\href {https://arxiv.org/abs/2211.01275} {\bibfield
  {journal} {\bibinfo  {journal} {arXiv:2211.01275}\ } (\bibinfo {year}
  {2022})}\BibitemShut {NoStop}%
\bibitem [{\citenamefont {Chen}\ \emph {et~al.}(2023)\citenamefont {Chen},
  \citenamefont {Bornet}, \citenamefont {Bintz}, \citenamefont {Emperauger},
  \citenamefont {Leclerc}, \citenamefont {Liu}, \citenamefont {Scholl},
  \citenamefont {Barredo}, \citenamefont {Hauschild}, \citenamefont
  {Chatterjee}, \citenamefont {Schuler}, \citenamefont {L\"{a}uchli},
  \citenamefont {Zaletel}, \citenamefont {Lahaye}, \citenamefont {Yao},\ and\
  \citenamefont {Browaeys}}]{Chen2023}%
  \BibitemOpen
  \bibfield  {author} {\bibinfo {author} {\bibfnamefont {Cheng}\ \bibnamefont
  {Chen}}, \bibinfo {author} {\bibfnamefont {Guillaume}\ \bibnamefont
  {Bornet}}, \bibinfo {author} {\bibfnamefont {Marcus}\ \bibnamefont {Bintz}},
  \bibinfo {author} {\bibfnamefont {Gabriel}\ \bibnamefont {Emperauger}},
  \bibinfo {author} {\bibfnamefont {Lucas}\ \bibnamefont {Leclerc}}, \bibinfo
  {author} {\bibfnamefont {Vincent~S.}\ \bibnamefont {Liu}}, \bibinfo {author}
  {\bibfnamefont {Pascal}\ \bibnamefont {Scholl}}, \bibinfo {author}
  {\bibfnamefont {Daniel}\ \bibnamefont {Barredo}}, \bibinfo {author}
  {\bibfnamefont {Johannes}\ \bibnamefont {Hauschild}}, \bibinfo {author}
  {\bibfnamefont {Shubhayu}\ \bibnamefont {Chatterjee}}, \bibinfo {author}
  {\bibfnamefont {Michael}\ \bibnamefont {Schuler}}, \bibinfo {author}
  {\bibfnamefont {Andreas~M.}\ \bibnamefont {L\"{a}uchli}}, \bibinfo {author}
  {\bibfnamefont {Michael~P.}\ \bibnamefont {Zaletel}}, \bibinfo {author}
  {\bibfnamefont {Thierry}\ \bibnamefont {Lahaye}}, \bibinfo {author}
  {\bibfnamefont {Norman~Y.}\ \bibnamefont {Yao}}, \ and\ \bibinfo {author}
  {\bibfnamefont {Antoine}\ \bibnamefont {Browaeys}},\ }\bibfield  {title}
  {\enquote {\bibinfo {title} {{Continuous symmetry breaking in a
  two-dimensional Rydberg array}},}\ }\href {\doibase
  10.1038/s41586-023-05859-2} {\bibfield  {journal} {\bibinfo  {journal}
  {Nature}\ }\textbf {\bibinfo {volume} {616}},\ \bibinfo {pages} {691--695}
  (\bibinfo {year} {2023})}\BibitemShut {NoStop}%
\bibitem [{\citenamefont {Tanzi}\ \emph {et~al.}(2019)\citenamefont {Tanzi},
  \citenamefont {Lucioni}, \citenamefont {Fam\`a}, \citenamefont {Catani},
  \citenamefont {Fioretti}, \citenamefont {Gabbanini}, \citenamefont {Bisset},
  \citenamefont {Santos},\ and\ \citenamefont
  {Modugno}}]{PhysRevLett.122.130405}%
  \BibitemOpen
  \bibfield  {author} {\bibinfo {author} {\bibfnamefont {L.}~\bibnamefont
  {Tanzi}}, \bibinfo {author} {\bibfnamefont {E.}~\bibnamefont {Lucioni}},
  \bibinfo {author} {\bibfnamefont {F.}~\bibnamefont {Fam\`a}}, \bibinfo
  {author} {\bibfnamefont {J.}~\bibnamefont {Catani}}, \bibinfo {author}
  {\bibfnamefont {A.}~\bibnamefont {Fioretti}}, \bibinfo {author}
  {\bibfnamefont {C.}~\bibnamefont {Gabbanini}}, \bibinfo {author}
  {\bibfnamefont {R.~N.}\ \bibnamefont {Bisset}}, \bibinfo {author}
  {\bibfnamefont {L.}~\bibnamefont {Santos}}, \ and\ \bibinfo {author}
  {\bibfnamefont {G.}~\bibnamefont {Modugno}},\ }\bibfield  {title} {\enquote
  {\bibinfo {title} {{Observation of a Dipolar Quantum Gas with Metastable
  Supersolid Properties}},}\ }\href {\doibase 10.1103/PhysRevLett.122.130405}
  {\bibfield  {journal} {\bibinfo  {journal} {Phys. Rev. Lett.}\ }\textbf
  {\bibinfo {volume} {122}},\ \bibinfo {pages} {130405} (\bibinfo {year}
  {2019})}\BibitemShut {NoStop}%
\bibitem [{\citenamefont {B\"ottcher}\ \emph {et~al.}(2019)\citenamefont
  {B\"ottcher}, \citenamefont {Schmidt}, \citenamefont {Wenzel}, \citenamefont
  {Hertkorn}, \citenamefont {Guo}, \citenamefont {Langen},\ and\ \citenamefont
  {Pfau}}]{PhysRevX.9.011051}%
  \BibitemOpen
  \bibfield  {author} {\bibinfo {author} {\bibfnamefont {Fabian}\ \bibnamefont
  {B\"ottcher}}, \bibinfo {author} {\bibfnamefont {Jan-Niklas}\ \bibnamefont
  {Schmidt}}, \bibinfo {author} {\bibfnamefont {Matthias}\ \bibnamefont
  {Wenzel}}, \bibinfo {author} {\bibfnamefont {Jens}\ \bibnamefont {Hertkorn}},
  \bibinfo {author} {\bibfnamefont {Mingyang}\ \bibnamefont {Guo}}, \bibinfo
  {author} {\bibfnamefont {Tim}\ \bibnamefont {Langen}}, \ and\ \bibinfo
  {author} {\bibfnamefont {Tilman}\ \bibnamefont {Pfau}},\ }\bibfield  {title}
  {\enquote {\bibinfo {title} {{Transient Supersolid Properties in an Array of
  Dipolar Quantum Droplets}},}\ }\href {\doibase 10.1103/PhysRevX.9.011051}
  {\bibfield  {journal} {\bibinfo  {journal} {Phys. Rev. X}\ }\textbf {\bibinfo
  {volume} {9}},\ \bibinfo {pages} {011051} (\bibinfo {year}
  {2019})}\BibitemShut {NoStop}%
\bibitem [{\citenamefont {Joshi}\ \emph {et~al.}(2022)\citenamefont {Joshi},
  \citenamefont {Kranzl}, \citenamefont {Schuckert}, \citenamefont {Lovas},
  \citenamefont {Maier}, \citenamefont {Blatt}, \citenamefont {Knap},\ and\
  \citenamefont {Roos}}]{Joshi2022}%
  \BibitemOpen
  \bibfield  {author} {\bibinfo {author} {\bibfnamefont {M.~K.}\ \bibnamefont
  {Joshi}}, \bibinfo {author} {\bibfnamefont {F.}~\bibnamefont {Kranzl}},
  \bibinfo {author} {\bibfnamefont {A.}~\bibnamefont {Schuckert}}, \bibinfo
  {author} {\bibfnamefont {I.}~\bibnamefont {Lovas}}, \bibinfo {author}
  {\bibfnamefont {C.}~\bibnamefont {Maier}}, \bibinfo {author} {\bibfnamefont
  {R.}~\bibnamefont {Blatt}}, \bibinfo {author} {\bibfnamefont
  {M.}~\bibnamefont {Knap}}, \ and\ \bibinfo {author} {\bibfnamefont {C.~F.}\
  \bibnamefont {Roos}},\ }\bibfield  {title} {\enquote {\bibinfo {title}
  {{Observing emergent hydrodynamics in a long-range quantum magnet}},}\ }\href
  {\doibase 10.1126/science.abk2400} {\bibfield  {journal} {\bibinfo  {journal}
  {Science}\ }\textbf {\bibinfo {volume} {376}},\ \bibinfo {pages} {720--724}
  (\bibinfo {year} {2022})}\BibitemShut {NoStop}%
\bibitem [{\citenamefont {Poulin}\ and\ \citenamefont
  {Wocjan}(2009)}]{Poulin2009}%
  \BibitemOpen
  \bibfield  {author} {\bibinfo {author} {\bibfnamefont {David}\ \bibnamefont
  {Poulin}}\ and\ \bibinfo {author} {\bibfnamefont {Pawel}\ \bibnamefont
  {Wocjan}},\ }\bibfield  {title} {\enquote {\bibinfo {title} {{Sampling from
  the Thermal Quantum Gibbs State and Evaluating Partition Functions with a
  Quantum Computer}},}\ }\href {\doibase 10.1103/PhysRevLett.103.220502}
  {\bibfield  {journal} {\bibinfo  {journal} {Phys. Rev. Lett.}\ }\textbf
  {\bibinfo {volume} {103}},\ \bibinfo {pages} {220502} (\bibinfo {year}
  {2009})}\BibitemShut {NoStop}%
\bibitem [{\citenamefont {Temme}\ \emph {et~al.}(2011)\citenamefont {Temme},
  \citenamefont {Osborne}, \citenamefont {Vollbrecht}, \citenamefont {Poulin},\
  and\ \citenamefont {Verstraete}}]{Temme2011}%
  \BibitemOpen
  \bibfield  {author} {\bibinfo {author} {\bibfnamefont {K.}~\bibnamefont
  {Temme}}, \bibinfo {author} {\bibfnamefont {T.~J.}\ \bibnamefont {Osborne}},
  \bibinfo {author} {\bibfnamefont {K.~G.}\ \bibnamefont {Vollbrecht}},
  \bibinfo {author} {\bibfnamefont {D.}~\bibnamefont {Poulin}}, \ and\ \bibinfo
  {author} {\bibfnamefont {F.}~\bibnamefont {Verstraete}},\ }\bibfield  {title}
  {\enquote {\bibinfo {title} {{Quantum Metropolis sampling}},}\ }\href
  {\doibase 10.1038/nature09770} {\bibfield  {journal} {\bibinfo  {journal}
  {Nature}\ }\textbf {\bibinfo {volume} {471}},\ \bibinfo {pages} {87--90}
  (\bibinfo {year} {2011})}\BibitemShut {NoStop}%
\bibitem [{\citenamefont {Chowdhury}\ \emph {et~al.}(2017)\citenamefont
  {Chowdhury}, ,\ and\ \citenamefont {Somma}}]{Chowdhury2017}%
  \BibitemOpen
  \bibfield  {author} {\bibinfo {author} {\bibfnamefont {Anirban~Narayan}\
  \bibnamefont {Chowdhury}}, , \ and\ \bibinfo {author} {\bibfnamefont
  {Rolando~D.}\ \bibnamefont {Somma}},\ }\bibfield  {title} {\enquote {\bibinfo
  {title} {{Quantum algorithms for Gibbs sampling and hitting-time
  estimation}},}\ }\href {\doibase 10.26421/qic17.1-2-3} {\bibfield  {journal}
  {\bibinfo  {journal} {Quantum Information and Computation}\ }\textbf
  {\bibinfo {volume} {17}},\ \bibinfo {pages} {41--64} (\bibinfo {year}
  {2017})}\BibitemShut {NoStop}%
\bibitem [{\citenamefont {Motta}\ \emph {et~al.}(2019)\citenamefont {Motta},
  \citenamefont {Sun}, \citenamefont {Tan}, \citenamefont {O'Rourke},
  \citenamefont {Ye}, \citenamefont {Minnich}, \citenamefont {Brand{\~{a}}o},\
  and\ \citenamefont {Chan}}]{Motta2019}%
  \BibitemOpen
  \bibfield  {author} {\bibinfo {author} {\bibfnamefont {Mario}\ \bibnamefont
  {Motta}}, \bibinfo {author} {\bibfnamefont {Chong}\ \bibnamefont {Sun}},
  \bibinfo {author} {\bibfnamefont {Adrian T.~K.}\ \bibnamefont {Tan}},
  \bibinfo {author} {\bibfnamefont {Matthew~J.}\ \bibnamefont {O'Rourke}},
  \bibinfo {author} {\bibfnamefont {Erika}\ \bibnamefont {Ye}}, \bibinfo
  {author} {\bibfnamefont {Austin~J.}\ \bibnamefont {Minnich}}, \bibinfo
  {author} {\bibfnamefont {Fernando G. S.~L.}\ \bibnamefont {Brand{\~{a}}o}}, \
  and\ \bibinfo {author} {\bibfnamefont {Garnet Kin-Lic}\ \bibnamefont
  {Chan}},\ }\bibfield  {title} {\enquote {\bibinfo {title} {Determining
  eigenstates and thermal states on a quantum computer using quantum imaginary
  time evolution},}\ }\href {\doibase 10.1038/s41567-019-0704-4} {\bibfield
  {journal} {\bibinfo  {journal} {Nature Physics}\ }\textbf {\bibinfo {volume}
  {16}},\ \bibinfo {pages} {205--210} (\bibinfo {year} {2019})}\BibitemShut
  {NoStop}%
\bibitem [{\citenamefont {Zhu}\ \emph {et~al.}(2020)\citenamefont {Zhu},
  \citenamefont {Johri}, \citenamefont {Linke}, \citenamefont {Landsman},
  \citenamefont {Alderete}, \citenamefont {Nguyen}, \citenamefont {Matsuura},
  \citenamefont {Hsieh},\ and\ \citenamefont {Monroe}}]{Zhu2020}%
  \BibitemOpen
  \bibfield  {author} {\bibinfo {author} {\bibfnamefont {D.}~\bibnamefont
  {Zhu}}, \bibinfo {author} {\bibfnamefont {S.}~\bibnamefont {Johri}}, \bibinfo
  {author} {\bibfnamefont {N.~M.}\ \bibnamefont {Linke}}, \bibinfo {author}
  {\bibfnamefont {K.~A.}\ \bibnamefont {Landsman}}, \bibinfo {author}
  {\bibfnamefont {C.~Huerta}\ \bibnamefont {Alderete}}, \bibinfo {author}
  {\bibfnamefont {N.~H.}\ \bibnamefont {Nguyen}}, \bibinfo {author}
  {\bibfnamefont {A.~Y.}\ \bibnamefont {Matsuura}}, \bibinfo {author}
  {\bibfnamefont {T.~H.}\ \bibnamefont {Hsieh}}, \ and\ \bibinfo {author}
  {\bibfnamefont {C.}~\bibnamefont {Monroe}},\ }\bibfield  {title} {\enquote
  {\bibinfo {title} {{Generation of thermofield double states and critical
  ground states with a quantum computer}},}\ }\href {\doibase
  10.1073/pnas.2006337117} {\bibfield  {journal} {\bibinfo  {journal}
  {Proceedings of the National Academy of Sciences}\ }\textbf {\bibinfo
  {volume} {117}},\ \bibinfo {pages} {25402--25406} (\bibinfo {year}
  {2020})}\BibitemShut {NoStop}%
\bibitem [{\citenamefont {Lu}\ \emph {et~al.}(2021)\citenamefont {Lu},
  \citenamefont {Ba\~nuls},\ and\ \citenamefont {Cirac}}]{Lu2021}%
  \BibitemOpen
  \bibfield  {author} {\bibinfo {author} {\bibfnamefont {Sirui}\ \bibnamefont
  {Lu}}, \bibinfo {author} {\bibfnamefont {Mari~Carmen}\ \bibnamefont
  {Ba\~nuls}}, \ and\ \bibinfo {author} {\bibfnamefont {J.~Ignacio}\
  \bibnamefont {Cirac}},\ }\bibfield  {title} {\enquote {\bibinfo {title}
  {{Algorithms for Quantum Simulation at Finite Energies}},}\ }\href {\doibase
  10.1103/PRXQuantum.2.020321} {\bibfield  {journal} {\bibinfo  {journal} {PRX
  Quantum}\ }\textbf {\bibinfo {volume} {2}},\ \bibinfo {pages} {020321}
  (\bibinfo {year} {2021})}\BibitemShut {NoStop}%
\bibitem [{\citenamefont {Shtanko}\ and\ \citenamefont
  {Movassagh}(2023)}]{shtanko2021}%
  \BibitemOpen
  \bibfield  {author} {\bibinfo {author} {\bibfnamefont {Oles}\ \bibnamefont
  {Shtanko}}\ and\ \bibinfo {author} {\bibfnamefont {Ramis}\ \bibnamefont
  {Movassagh}},\ }\bibfield  {title} {\enquote {\bibinfo {title} {Preparing
  thermal states on noiseless and noisy programmable quantum processors},}\
  }\href {https://doi.org/10.48550/arXiv.2112.14688} {\bibfield  {journal}
  {\bibinfo  {journal} {arXiv:2112.14688}\ } (\bibinfo {year}
  {2023})}\BibitemShut {NoStop}%
\bibitem [{\citenamefont {Tazhigulov}\ \emph {et~al.}(2022)\citenamefont
  {Tazhigulov}, \citenamefont {Sun}, \citenamefont {Haghshenas}, \citenamefont
  {Zhai}, \citenamefont {Tan}, \citenamefont {Rubin}, \citenamefont {Babbush},
  \citenamefont {Minnich},\ and\ \citenamefont {Chan}}]{tazhigulov2022}%
  \BibitemOpen
  \bibfield  {author} {\bibinfo {author} {\bibfnamefont {Ruslan~N.}\
  \bibnamefont {Tazhigulov}}, \bibinfo {author} {\bibfnamefont {Shi-Ning}\
  \bibnamefont {Sun}}, \bibinfo {author} {\bibfnamefont {Reza}\ \bibnamefont
  {Haghshenas}}, \bibinfo {author} {\bibfnamefont {Huanchen}\ \bibnamefont
  {Zhai}}, \bibinfo {author} {\bibfnamefont {Adrian~T.K.}\ \bibnamefont {Tan}},
  \bibinfo {author} {\bibfnamefont {Nicholas~C.}\ \bibnamefont {Rubin}},
  \bibinfo {author} {\bibfnamefont {Ryan}\ \bibnamefont {Babbush}}, \bibinfo
  {author} {\bibfnamefont {Austin~J.}\ \bibnamefont {Minnich}}, \ and\ \bibinfo
  {author} {\bibfnamefont {Garnet Kin-Lic}\ \bibnamefont {Chan}},\ }\bibfield
  {title} {\enquote {\bibinfo {title} {{Simulating Models of Challenging
  Correlated Molecules and Materials on the Sycamore Quantum Processor}},}\
  }\href {\doibase 10.1103/PRXQuantum.3.040318} {\bibfield  {journal} {\bibinfo
   {journal} {PRX Quantum}\ }\textbf {\bibinfo {volume} {3}},\ \bibinfo {pages}
  {040318} (\bibinfo {year} {2022})}\BibitemShut {NoStop}%
\bibitem [{\citenamefont {Schuckert}\ \emph {et~al.}(2023)\citenamefont
  {Schuckert}, \citenamefont {Bohrdt}, \citenamefont {Crane},\ and\
  \citenamefont {Knap}}]{Schuckert2022}%
  \BibitemOpen
  \bibfield  {author} {\bibinfo {author} {\bibfnamefont {Alexander}\
  \bibnamefont {Schuckert}}, \bibinfo {author} {\bibfnamefont {Annabelle}\
  \bibnamefont {Bohrdt}}, \bibinfo {author} {\bibfnamefont {Eleanor}\
  \bibnamefont {Crane}}, \ and\ \bibinfo {author} {\bibfnamefont {Michael}\
  \bibnamefont {Knap}},\ }\bibfield  {title} {\enquote {\bibinfo {title}
  {Probing finite-temperature observables in quantum simulators of spin systems
  with short-time dynamics},}\ }\href {\doibase 10.1103/PhysRevB.107.L140410}
  {\bibfield  {journal} {\bibinfo  {journal} {Phys. Rev. B}\ }\textbf {\bibinfo
  {volume} {107}},\ \bibinfo {pages} {L140410} (\bibinfo {year}
  {2023})}\BibitemShut {NoStop}%
\bibitem [{\citenamefont {Hémery}\ \emph {et~al.}(2023)\citenamefont
  {Hémery}, \citenamefont {Ghanem}, \citenamefont {Crane}, \citenamefont
  {Campbell}, \citenamefont {Dreiling}, \citenamefont {Figgatt}, \citenamefont
  {Foltz}, \citenamefont {Gaebler}, \citenamefont {Johansen}, \citenamefont
  {Mills}, \citenamefont {Moses}, \citenamefont {Pino}, \citenamefont
  {Ransford}, \citenamefont {Rowe}, \citenamefont {Siegfried}, \citenamefont
  {Stutz}, \citenamefont {Dreyer}, \citenamefont {Schuckert},\ and\
  \citenamefont {Nigmatullin}}]{hemery2023}%
  \BibitemOpen
  \bibfield  {author} {\bibinfo {author} {\bibfnamefont {Kévin}\ \bibnamefont
  {Hémery}}, \bibinfo {author} {\bibfnamefont {Khaldoon}\ \bibnamefont
  {Ghanem}}, \bibinfo {author} {\bibfnamefont {Eleanor}\ \bibnamefont {Crane}},
  \bibinfo {author} {\bibfnamefont {Sara~L.}\ \bibnamefont {Campbell}},
  \bibinfo {author} {\bibfnamefont {Joan~M.}\ \bibnamefont {Dreiling}},
  \bibinfo {author} {\bibfnamefont {Caroline}\ \bibnamefont {Figgatt}},
  \bibinfo {author} {\bibfnamefont {Cameron}\ \bibnamefont {Foltz}}, \bibinfo
  {author} {\bibfnamefont {John~P.}\ \bibnamefont {Gaebler}}, \bibinfo {author}
  {\bibfnamefont {Jacob}\ \bibnamefont {Johansen}}, \bibinfo {author}
  {\bibfnamefont {Michael}\ \bibnamefont {Mills}}, \bibinfo {author}
  {\bibfnamefont {Steven~A.}\ \bibnamefont {Moses}}, \bibinfo {author}
  {\bibfnamefont {Juan~M.}\ \bibnamefont {Pino}}, \bibinfo {author}
  {\bibfnamefont {Anthony}\ \bibnamefont {Ransford}}, \bibinfo {author}
  {\bibfnamefont {Mary}\ \bibnamefont {Rowe}}, \bibinfo {author} {\bibfnamefont
  {Peter}\ \bibnamefont {Siegfried}}, \bibinfo {author} {\bibfnamefont
  {Russell~P.}\ \bibnamefont {Stutz}}, \bibinfo {author} {\bibfnamefont
  {Henrik}\ \bibnamefont {Dreyer}}, \bibinfo {author} {\bibfnamefont
  {Alexander}\ \bibnamefont {Schuckert}}, \ and\ \bibinfo {author}
  {\bibfnamefont {Ramil}\ \bibnamefont {Nigmatullin}},\ }\bibfield  {title}
  {\enquote {\bibinfo {title} {{Measuring the Loschmidt amplitude for
  finite-energy properties of the Fermi-Hubbard model on an ion-trap quantum
  computer}},}\ }\href {https://doi.org/10.48550/arXiv.2309.10552} {\bibfield
  {journal} {\bibinfo  {journal} {arXiv:2309.10552}\ } (\bibinfo {year}
  {2023})}\BibitemShut {NoStop}%
\bibitem [{\citenamefont {Srednicki}(1996)}]{Srednicki1996}%
  \BibitemOpen
  \bibfield  {author} {\bibinfo {author} {\bibfnamefont {Mark}\ \bibnamefont
  {Srednicki}},\ }\bibfield  {title} {\enquote {\bibinfo {title} {Thermal
  fluctuations in quantized chaotic systems},}\ }\href {\doibase
  10.1088/0305-4470/29/4/003} {\bibfield  {journal} {\bibinfo  {journal}
  {Journal of Physics A: Mathematical and General}\ }\textbf {\bibinfo {volume}
  {29}},\ \bibinfo {pages} {L75--L79} (\bibinfo {year} {1996})}\BibitemShut
  {NoStop}%
\bibitem [{\citenamefont {Rigol}\ \emph {et~al.}(2008)\citenamefont {Rigol},
  \citenamefont {Dunjko},\ and\ \citenamefont {Olshanii}}]{Rigol2008}%
  \BibitemOpen
  \bibfield  {author} {\bibinfo {author} {\bibfnamefont {Marcos}\ \bibnamefont
  {Rigol}}, \bibinfo {author} {\bibfnamefont {Vanja}\ \bibnamefont {Dunjko}}, \
  and\ \bibinfo {author} {\bibfnamefont {Maxim}\ \bibnamefont {Olshanii}},\
  }\bibfield  {title} {\enquote {\bibinfo {title} {{Thermalization and its
  mechanism for generic isolated quantum systems}},}\ }\href {\doibase
  10.1038/nature06838} {\bibfield  {journal} {\bibinfo  {journal} {Nature}\
  }\textbf {\bibinfo {volume} {452}},\ \bibinfo {pages} {854--858} (\bibinfo
  {year} {2008})}\BibitemShut {NoStop}%
\bibitem [{\citenamefont {Trotzky}\ \emph {et~al.}(2012)\citenamefont
  {Trotzky}, \citenamefont {Chen}, \citenamefont {Flesch}, \citenamefont
  {McCulloch}, \citenamefont {Schollw\"{o}ck}, \citenamefont {Eisert},\ and\
  \citenamefont {Bloch}}]{Trotzky2012}%
  \BibitemOpen
  \bibfield  {author} {\bibinfo {author} {\bibfnamefont {S.}~\bibnamefont
  {Trotzky}}, \bibinfo {author} {\bibfnamefont {Y-A.}\ \bibnamefont {Chen}},
  \bibinfo {author} {\bibfnamefont {A.}~\bibnamefont {Flesch}}, \bibinfo
  {author} {\bibfnamefont {I.~P.}\ \bibnamefont {McCulloch}}, \bibinfo {author}
  {\bibfnamefont {U.}~\bibnamefont {Schollw\"{o}ck}}, \bibinfo {author}
  {\bibfnamefont {J.}~\bibnamefont {Eisert}}, \ and\ \bibinfo {author}
  {\bibfnamefont {I.}~\bibnamefont {Bloch}},\ }\bibfield  {title} {\enquote
  {\bibinfo {title} {Probing the relaxation towards equilibrium in an isolated
  strongly correlated one-dimensional bose~gas},}\ }\href {\doibase
  10.1038/nphys2232} {\bibfield  {journal} {\bibinfo  {journal} {Nature
  Physics}\ }\textbf {\bibinfo {volume} {8}},\ \bibinfo {pages} {325--330}
  (\bibinfo {year} {2012})}\BibitemShut {NoStop}%
\bibitem [{\citenamefont {Kaufman}\ \emph {et~al.}(2016)\citenamefont
  {Kaufman}, \citenamefont {Tai}, \citenamefont {Lukin}, \citenamefont
  {Rispoli}, \citenamefont {Schittko}, \citenamefont {Preiss},\ and\
  \citenamefont {Greiner}}]{Kaufman2016}%
  \BibitemOpen
  \bibfield  {author} {\bibinfo {author} {\bibfnamefont {Adam~M.}\ \bibnamefont
  {Kaufman}}, \bibinfo {author} {\bibfnamefont {M.~Eric}\ \bibnamefont {Tai}},
  \bibinfo {author} {\bibfnamefont {Alexander}\ \bibnamefont {Lukin}}, \bibinfo
  {author} {\bibfnamefont {Matthew}\ \bibnamefont {Rispoli}}, \bibinfo {author}
  {\bibfnamefont {Robert}\ \bibnamefont {Schittko}}, \bibinfo {author}
  {\bibfnamefont {Philipp~M.}\ \bibnamefont {Preiss}}, \ and\ \bibinfo {author}
  {\bibfnamefont {Markus}\ \bibnamefont {Greiner}},\ }\bibfield  {title}
  {\enquote {\bibinfo {title} {Quantum thermalization through entanglement in
  an isolated many-body system},}\ }\href {\doibase 10.1126/science.aaf6725}
  {\bibfield  {journal} {\bibinfo  {journal} {Science}\ }\textbf {\bibinfo
  {volume} {353}},\ \bibinfo {pages} {794--800} (\bibinfo {year}
  {2016})}\BibitemShut {NoStop}%
\bibitem [{sup()}]{supp}%
  \BibitemOpen
  \href@noop {} {}\bibinfo {note} {Materials and methods are available as
  supplementary materials.}\BibitemShut {Stop}%
\bibitem [{\citenamefont {Katz}\ \emph {et~al.}(2023)\citenamefont {Katz},
  \citenamefont {Feng}, \citenamefont {Risinger}, \citenamefont {Monroe},\ and\
  \citenamefont {Cetina}}]{katz2023}%
  \BibitemOpen
  \bibfield  {author} {\bibinfo {author} {\bibfnamefont {Or}~\bibnamefont
  {Katz}}, \bibinfo {author} {\bibfnamefont {Lei}\ \bibnamefont {Feng}},
  \bibinfo {author} {\bibfnamefont {Andrew}\ \bibnamefont {Risinger}}, \bibinfo
  {author} {\bibfnamefont {Christopher}\ \bibnamefont {Monroe}}, \ and\
  \bibinfo {author} {\bibfnamefont {Marko}\ \bibnamefont {Cetina}},\ }\bibfield
   {title} {\enquote {\bibinfo {title} {Demonstration of three- and four-body
  interactions between trapped-ion spins},}\ }\href {\doibase
  10.1038/s41567-023-02102-7} {\bibfield  {journal} {\bibinfo  {journal}
  {Nature Physics}\ }\textbf {\bibinfo {volume} {19}},\ \bibinfo {pages}
  {1452--1458} (\bibinfo {year} {2023})}\BibitemShut {NoStop}%
\bibitem [{\citenamefont {Porras}\ and\ \citenamefont
  {Cirac}(2004)}]{Porras2004}%
  \BibitemOpen
  \bibfield  {author} {\bibinfo {author} {\bibfnamefont {D.}~\bibnamefont
  {Porras}}\ and\ \bibinfo {author} {\bibfnamefont {J.~I.}\ \bibnamefont
  {Cirac}},\ }\bibfield  {title} {\enquote {\bibinfo {title} {{Effective
  Quantum Spin Systems with Trapped Ions}},}\ }\href
  {https://doi.org/10.1103/physrevlett.92.207901} {\bibfield  {journal}
  {\bibinfo  {journal} {Phys. Rev. Lett.}\ }\textbf {\bibinfo {volume} {92}},\
  \bibinfo {pages} {207901} (\bibinfo {year} {2004})}\BibitemShut {NoStop}%
\bibitem [{\citenamefont {Monroe}\ \emph {et~al.}(2021)\citenamefont {Monroe},
  \citenamefont {Campbell}, \citenamefont {Duan}, \citenamefont {Gong},
  \citenamefont {Gorshkov}, \citenamefont {Hess}, \citenamefont {Islam},
  \citenamefont {Kim}, \citenamefont {Linke}, \citenamefont {Pagano},
  \citenamefont {Richerme}, \citenamefont {Senko},\ and\ \citenamefont
  {Yao}}]{Monroe2022}%
  \BibitemOpen
  \bibfield  {author} {\bibinfo {author} {\bibfnamefont {C.}~\bibnamefont
  {Monroe}}, \bibinfo {author} {\bibfnamefont {W.~C.}\ \bibnamefont
  {Campbell}}, \bibinfo {author} {\bibfnamefont {L.-M.}\ \bibnamefont {Duan}},
  \bibinfo {author} {\bibfnamefont {Z.-X.}\ \bibnamefont {Gong}}, \bibinfo
  {author} {\bibfnamefont {A.~V.}\ \bibnamefont {Gorshkov}}, \bibinfo {author}
  {\bibfnamefont {P.~W.}\ \bibnamefont {Hess}}, \bibinfo {author}
  {\bibfnamefont {R.}~\bibnamefont {Islam}}, \bibinfo {author} {\bibfnamefont
  {K.}~\bibnamefont {Kim}}, \bibinfo {author} {\bibfnamefont {N.~M.}\
  \bibnamefont {Linke}}, \bibinfo {author} {\bibfnamefont {G.}~\bibnamefont
  {Pagano}}, \bibinfo {author} {\bibfnamefont {P.}~\bibnamefont {Richerme}},
  \bibinfo {author} {\bibfnamefont {C.}~\bibnamefont {Senko}}, \ and\ \bibinfo
  {author} {\bibfnamefont {N.~Y.}\ \bibnamefont {Yao}},\ }\bibfield  {title}
  {\enquote {\bibinfo {title} {Programmable quantum simulations of spin systems
  with trapped ions},}\ }\href {\doibase 10.1103/RevModPhys.93.025001}
  {\bibfield  {journal} {\bibinfo  {journal} {Rev. Mod. Phys.}\ }\textbf
  {\bibinfo {volume} {93}},\ \bibinfo {pages} {025001} (\bibinfo {year}
  {2021})}\BibitemShut {NoStop}%
\bibitem [{\citenamefont {Kac}\ \emph {et~al.}(1963)\citenamefont {Kac},
  \citenamefont {Uhlenbeck},\ and\ \citenamefont {Hemmer}}]{Kac1963}%
  \BibitemOpen
  \bibfield  {author} {\bibinfo {author} {\bibfnamefont {M.}~\bibnamefont
  {Kac}}, \bibinfo {author} {\bibfnamefont {G.~E.}\ \bibnamefont {Uhlenbeck}},
  \ and\ \bibinfo {author} {\bibfnamefont {P.~C.}\ \bibnamefont {Hemmer}},\
  }\bibfield  {title} {\enquote {\bibinfo {title} {{On the van der Waals Theory
  of the Vapor-Liquid Equilibrium. I. Discussion of a One-Dimensional
  Model}},}\ }\href {\doibase 10.1063/1.1703946} {\bibfield  {journal}
  {\bibinfo  {journal} {Journal of Mathematical Physics}\ }\textbf {\bibinfo
  {volume} {4}},\ \bibinfo {pages} {216--228} (\bibinfo {year}
  {1963})}\BibitemShut {NoStop}%
\bibitem [{\citenamefont {Yang}\ \emph {et~al.}(2023)\citenamefont {Yang},
  \citenamefont {Christianen}, \citenamefont {Coll-Vinent}, \citenamefont
  {Smelyanskiy}, \citenamefont {Ba\~nuls}, \citenamefont {O'Brien},
  \citenamefont {Wild},\ and\ \citenamefont {Cirac}}]{Yang2023}%
  \BibitemOpen
  \bibfield  {author} {\bibinfo {author} {\bibfnamefont {Yilun}\ \bibnamefont
  {Yang}}, \bibinfo {author} {\bibfnamefont {Arthur}\ \bibnamefont
  {Christianen}}, \bibinfo {author} {\bibfnamefont {Sandra}\ \bibnamefont
  {Coll-Vinent}}, \bibinfo {author} {\bibfnamefont {Vadim}\ \bibnamefont
  {Smelyanskiy}}, \bibinfo {author} {\bibfnamefont {Mari~Carmen}\ \bibnamefont
  {Ba\~nuls}}, \bibinfo {author} {\bibfnamefont {Thomas~E.}\ \bibnamefont
  {O'Brien}}, \bibinfo {author} {\bibfnamefont {Dominik~S.}\ \bibnamefont
  {Wild}}, \ and\ \bibinfo {author} {\bibfnamefont {J.~Ignacio}\ \bibnamefont
  {Cirac}},\ }\bibfield  {title} {\enquote {\bibinfo {title} {{Simulating
  Prethermalization Using Near-Term Quantum Computers}},}\ }\href {\doibase
  10.1103/PRXQuantum.4.030320} {\bibfield  {journal} {\bibinfo  {journal} {PRX
  Quantum}\ }\textbf {\bibinfo {volume} {4}},\ \bibinfo {pages} {030320}
  (\bibinfo {year} {2023})}\BibitemShut {NoStop}%
\bibitem [{\citenamefont {Gong}\ \emph {et~al.}(2016)\citenamefont {Gong},
  \citenamefont {Maghrebi}, \citenamefont {Hu}, \citenamefont {Foss-Feig},
  \citenamefont {Richerme}, \citenamefont {Monroe},\ and\ \citenamefont
  {Gorshkov}}]{gong2016kaleidoscope}%
  \BibitemOpen
  \bibfield  {author} {\bibinfo {author} {\bibfnamefont {Z-X}\ \bibnamefont
  {Gong}}, \bibinfo {author} {\bibfnamefont {Mohammad~F}\ \bibnamefont
  {Maghrebi}}, \bibinfo {author} {\bibfnamefont {Anzi}\ \bibnamefont {Hu}},
  \bibinfo {author} {\bibfnamefont {Michael}\ \bibnamefont {Foss-Feig}},
  \bibinfo {author} {\bibfnamefont {Phillip}\ \bibnamefont {Richerme}},
  \bibinfo {author} {\bibfnamefont {Christopher}\ \bibnamefont {Monroe}}, \
  and\ \bibinfo {author} {\bibfnamefont {Alexey~V}\ \bibnamefont {Gorshkov}},\
  }\bibfield  {title} {\enquote {\bibinfo {title} {Kaleidoscope of quantum
  phases in a long-range interacting spin-1 chain},}\ }\href@noop {} {\bibfield
   {journal} {\bibinfo  {journal} {Physical Review B}\ }\textbf {\bibinfo
  {volume} {93}},\ \bibinfo {pages} {205115} (\bibinfo {year}
  {2016})}\BibitemShut {NoStop}%
\bibitem [{\citenamefont {Defenu}\ \emph
  {et~al.}(2023{\natexlab{b}})\citenamefont {Defenu}, \citenamefont {Donner},
  \citenamefont {Macr{\`\i}}, \citenamefont {Pagano}, \citenamefont {Ruffo},\
  and\ \citenamefont {Trombettoni}}]{defenu2023long}%
  \BibitemOpen
  \bibfield  {author} {\bibinfo {author} {\bibfnamefont {Nicol{\`o}}\
  \bibnamefont {Defenu}}, \bibinfo {author} {\bibfnamefont {Tobias}\
  \bibnamefont {Donner}}, \bibinfo {author} {\bibfnamefont {Tommaso}\
  \bibnamefont {Macr{\`\i}}}, \bibinfo {author} {\bibfnamefont {Guido}\
  \bibnamefont {Pagano}}, \bibinfo {author} {\bibfnamefont {Stefano}\
  \bibnamefont {Ruffo}}, \ and\ \bibinfo {author} {\bibfnamefont {Andrea}\
  \bibnamefont {Trombettoni}},\ }\bibfield  {title} {\enquote {\bibinfo {title}
  {{Long-range interacting quantum systems}},}\ }\href@noop {} {\bibfield
  {journal} {\bibinfo  {journal} {Reviews of Modern Physics}\ }\textbf
  {\bibinfo {volume} {95}},\ \bibinfo {pages} {035002} (\bibinfo {year}
  {2023}{\natexlab{b}})}\BibitemShut {NoStop}%
\bibitem [{\citenamefont {Hartke}\ \emph {et~al.}(2020)\citenamefont {Hartke},
  \citenamefont {Oreg}, \citenamefont {Jia},\ and\ \citenamefont
  {Zwierlein}}]{Hartke2020}%
  \BibitemOpen
  \bibfield  {author} {\bibinfo {author} {\bibfnamefont {Thomas}\ \bibnamefont
  {Hartke}}, \bibinfo {author} {\bibfnamefont {Botond}\ \bibnamefont {Oreg}},
  \bibinfo {author} {\bibfnamefont {Ningyuan}\ \bibnamefont {Jia}}, \ and\
  \bibinfo {author} {\bibfnamefont {Martin}\ \bibnamefont {Zwierlein}},\
  }\bibfield  {title} {\enquote {\bibinfo {title} {{Doublon-Hole Correlations
  and Fluctuation Thermometry in a Fermi-Hubbard Gas}},}\ }\href
  {https://doi.org/10.1103/physrevlett.125.113601} {\bibfield  {journal}
  {\bibinfo  {journal} {Phys. Rev. Lett.}\ }\textbf {\bibinfo {volume} {125}},\
  \bibinfo {pages} {113601} (\bibinfo {year} {2020})}\BibitemShut {NoStop}%
\bibitem [{\citenamefont {Schuckert}\ and\ \citenamefont
  {Knap}(2020)}]{Schuckert2020}%
  \BibitemOpen
  \bibfield  {author} {\bibinfo {author} {\bibfnamefont {Alexander}\
  \bibnamefont {Schuckert}}\ and\ \bibinfo {author} {\bibfnamefont {Michael}\
  \bibnamefont {Knap}},\ }\bibfield  {title} {\enquote {\bibinfo {title}
  {Probing eigenstate thermalization in quantum simulators via
  fluctuation-dissipation relations},}\ }\href
  {https://doi.org/10.1103/physrevresearch.2.043315} {\bibfield  {journal}
  {\bibinfo  {journal} {Phys. Rev. Res.}\ }\textbf {\bibinfo {volume} {2}},\
  \bibinfo {pages} {043315} (\bibinfo {year} {2020})}\BibitemShut {NoStop}%
\bibitem [{\citenamefont {Semeghini}\ \emph {et~al.}(2021)\citenamefont
  {Semeghini}, \citenamefont {Levine}, \citenamefont {Keesling}, \citenamefont
  {Ebadi}, \citenamefont {Wang}, \citenamefont {Bluvstein}, \citenamefont
  {Verresen}, \citenamefont {Pichler}, \citenamefont {Kalinowski},
  \citenamefont {Samajdar}, \citenamefont {Omran}, \citenamefont {Sachdev},
  \citenamefont {Vishwanath}, \citenamefont {Greiner}, \citenamefont
  {Vuleti{\'{c}}},\ and\ \citenamefont {Lukin}}]{Semeghini2021}%
  \BibitemOpen
  \bibfield  {author} {\bibinfo {author} {\bibfnamefont {G.}~\bibnamefont
  {Semeghini}}, \bibinfo {author} {\bibfnamefont {H.}~\bibnamefont {Levine}},
  \bibinfo {author} {\bibfnamefont {A.}~\bibnamefont {Keesling}}, \bibinfo
  {author} {\bibfnamefont {S.}~\bibnamefont {Ebadi}}, \bibinfo {author}
  {\bibfnamefont {T.~T.}\ \bibnamefont {Wang}}, \bibinfo {author}
  {\bibfnamefont {D.}~\bibnamefont {Bluvstein}}, \bibinfo {author}
  {\bibfnamefont {R.}~\bibnamefont {Verresen}}, \bibinfo {author}
  {\bibfnamefont {H.}~\bibnamefont {Pichler}}, \bibinfo {author} {\bibfnamefont
  {M.}~\bibnamefont {Kalinowski}}, \bibinfo {author} {\bibfnamefont
  {R.}~\bibnamefont {Samajdar}}, \bibinfo {author} {\bibfnamefont
  {A.}~\bibnamefont {Omran}}, \bibinfo {author} {\bibfnamefont
  {S.}~\bibnamefont {Sachdev}}, \bibinfo {author} {\bibfnamefont
  {A.}~\bibnamefont {Vishwanath}}, \bibinfo {author} {\bibfnamefont
  {M.}~\bibnamefont {Greiner}}, \bibinfo {author} {\bibfnamefont
  {V.}~\bibnamefont {Vuleti{\'{c}}}}, \ and\ \bibinfo {author} {\bibfnamefont
  {M.~D.}\ \bibnamefont {Lukin}},\ }\bibfield  {title} {\enquote {\bibinfo
  {title} {Probing topological spin liquids on a programmable quantum
  simulator},}\ }\href {\doibase 10.1126/science.abi8794} {\bibfield  {journal}
  {\bibinfo  {journal} {Science}\ }\textbf {\bibinfo {volume} {374}},\ \bibinfo
  {pages} {1242--1247} (\bibinfo {year} {2021})}\BibitemShut {NoStop}%
\bibitem [{\citenamefont {Satzinger}\ \emph {et~al.}(2021)\citenamefont
  {Satzinger}, \citenamefont {Liu}, \citenamefont {Smith}, \citenamefont
  {Knapp}, \citenamefont {Newman}, \citenamefont {Jones}, \citenamefont {Chen},
  \citenamefont {Quintana}, \citenamefont {Mi}, \citenamefont {Dunsworth},
  \citenamefont {Gidney}, \citenamefont {Aleiner}, \citenamefont {Arute},
  \citenamefont {Arya}, \citenamefont {Atalaya}, \citenamefont {Babbush},
  \citenamefont {Bardin}, \citenamefont {Barends}, \citenamefont {Basso},
  \citenamefont {Bengtsson}, \citenamefont {Bilmes}, \citenamefont {Broughton},
  \citenamefont {Buckley}, \citenamefont {Buell}, \citenamefont {Burkett},
  \citenamefont {Bushnell}, \citenamefont {Chiaro}, \citenamefont {Collins},
  \citenamefont {Courtney}, \citenamefont {Demura}, \citenamefont {Derk},
  \citenamefont {Eppens}, \citenamefont {Erickson}, \citenamefont {Faoro},
  \citenamefont {Farhi}, \citenamefont {Fowler}, \citenamefont {Foxen},
  \citenamefont {Giustina}, \citenamefont {Greene}, \citenamefont {Gross},
  \citenamefont {Harrigan}, \citenamefont {Harrington}, \citenamefont {Hilton},
  \citenamefont {Hong}, \citenamefont {Huang}, \citenamefont {Huggins},
  \citenamefont {Ioffe}, \citenamefont {Isakov}, \citenamefont {Jeffrey},
  \citenamefont {Jiang}, \citenamefont {Kafri}, \citenamefont {Kechedzhi},
  \citenamefont {Khattar}, \citenamefont {Kim}, \citenamefont {Klimov},
  \citenamefont {Korotkov}, \citenamefont {Kostritsa}, \citenamefont
  {Landhuis}, \citenamefont {Laptev}, \citenamefont {Locharla}, \citenamefont
  {Lucero}, \citenamefont {Martin}, \citenamefont {McClean}, \citenamefont
  {McEwen}, \citenamefont {Miao}, \citenamefont {Mohseni}, \citenamefont
  {Montazeri}, \citenamefont {Mruczkiewicz}, \citenamefont {Mutus},
  \citenamefont {Naaman}, \citenamefont {Neeley}, \citenamefont {Neill},
  \citenamefont {Niu}, \citenamefont {O'Brien}, \citenamefont {Opremcak},
  \citenamefont {Pat{\'{o}}}, \citenamefont {Petukhov}, \citenamefont {Rubin},
  \citenamefont {Sank}, \citenamefont {Shvarts}, \citenamefont {Strain},
  \citenamefont {Szalay}, \citenamefont {Villalonga}, \citenamefont {White},
  \citenamefont {Yao}, \citenamefont {Yeh}, \citenamefont {Yoo}, \citenamefont
  {Zalcman}, \citenamefont {Neven}, \citenamefont {Boixo}, \citenamefont
  {Megrant}, \citenamefont {Chen}, \citenamefont {Kelly}, \citenamefont
  {Smelyanskiy}, \citenamefont {Kitaev}, \citenamefont {Knap}, \citenamefont
  {Pollmann},\ and\ \citenamefont {Roushan}}]{Satzinger2021}%
  \BibitemOpen
  \bibfield  {author} {\bibinfo {author} {\bibfnamefont {K.~J.}\ \bibnamefont
  {Satzinger}}, \bibinfo {author} {\bibfnamefont {Y.-J}\ \bibnamefont {Liu}},
  \bibinfo {author} {\bibfnamefont {A.}~\bibnamefont {Smith}}, \bibinfo
  {author} {\bibfnamefont {C.}~\bibnamefont {Knapp}}, \bibinfo {author}
  {\bibfnamefont {M.}~\bibnamefont {Newman}}, \bibinfo {author} {\bibfnamefont
  {C.}~\bibnamefont {Jones}}, \bibinfo {author} {\bibfnamefont
  {Z.}~\bibnamefont {Chen}}, \bibinfo {author} {\bibfnamefont {C.}~\bibnamefont
  {Quintana}}, \bibinfo {author} {\bibfnamefont {X.}~\bibnamefont {Mi}},
  \bibinfo {author} {\bibfnamefont {A.}~\bibnamefont {Dunsworth}}, \bibinfo
  {author} {\bibfnamefont {C.}~\bibnamefont {Gidney}}, \bibinfo {author}
  {\bibfnamefont {I.}~\bibnamefont {Aleiner}}, \bibinfo {author} {\bibfnamefont
  {F.}~\bibnamefont {Arute}}, \bibinfo {author} {\bibfnamefont
  {K.}~\bibnamefont {Arya}}, \bibinfo {author} {\bibfnamefont {J.}~\bibnamefont
  {Atalaya}}, \bibinfo {author} {\bibfnamefont {R.}~\bibnamefont {Babbush}},
  \bibinfo {author} {\bibfnamefont {J.~C.}\ \bibnamefont {Bardin}}, \bibinfo
  {author} {\bibfnamefont {R.}~\bibnamefont {Barends}}, \bibinfo {author}
  {\bibfnamefont {J.}~\bibnamefont {Basso}}, \bibinfo {author} {\bibfnamefont
  {A.}~\bibnamefont {Bengtsson}}, \bibinfo {author} {\bibfnamefont
  {A.}~\bibnamefont {Bilmes}}, \bibinfo {author} {\bibfnamefont
  {M.}~\bibnamefont {Broughton}}, \bibinfo {author} {\bibfnamefont {B.~B.}\
  \bibnamefont {Buckley}}, \bibinfo {author} {\bibfnamefont {D.~A.}\
  \bibnamefont {Buell}}, \bibinfo {author} {\bibfnamefont {B.}~\bibnamefont
  {Burkett}}, \bibinfo {author} {\bibfnamefont {N.}~\bibnamefont {Bushnell}},
  \bibinfo {author} {\bibfnamefont {B.}~\bibnamefont {Chiaro}}, \bibinfo
  {author} {\bibfnamefont {R.}~\bibnamefont {Collins}}, \bibinfo {author}
  {\bibfnamefont {W.}~\bibnamefont {Courtney}}, \bibinfo {author}
  {\bibfnamefont {S.}~\bibnamefont {Demura}}, \bibinfo {author} {\bibfnamefont
  {A.~R.}\ \bibnamefont {Derk}}, \bibinfo {author} {\bibfnamefont
  {D.}~\bibnamefont {Eppens}}, \bibinfo {author} {\bibfnamefont
  {C.}~\bibnamefont {Erickson}}, \bibinfo {author} {\bibfnamefont
  {L.}~\bibnamefont {Faoro}}, \bibinfo {author} {\bibfnamefont
  {E.}~\bibnamefont {Farhi}}, \bibinfo {author} {\bibfnamefont {A.~G.}\
  \bibnamefont {Fowler}}, \bibinfo {author} {\bibfnamefont {B.}~\bibnamefont
  {Foxen}}, \bibinfo {author} {\bibfnamefont {M.}~\bibnamefont {Giustina}},
  \bibinfo {author} {\bibfnamefont {A.}~\bibnamefont {Greene}}, \bibinfo
  {author} {\bibfnamefont {J.~A.}\ \bibnamefont {Gross}}, \bibinfo {author}
  {\bibfnamefont {M.~P.}\ \bibnamefont {Harrigan}}, \bibinfo {author}
  {\bibfnamefont {S.~D.}\ \bibnamefont {Harrington}}, \bibinfo {author}
  {\bibfnamefont {J.}~\bibnamefont {Hilton}}, \bibinfo {author} {\bibfnamefont
  {S.}~\bibnamefont {Hong}}, \bibinfo {author} {\bibfnamefont {T.}~\bibnamefont
  {Huang}}, \bibinfo {author} {\bibfnamefont {W.~J.}\ \bibnamefont {Huggins}},
  \bibinfo {author} {\bibfnamefont {L.~B.}\ \bibnamefont {Ioffe}}, \bibinfo
  {author} {\bibfnamefont {S.~V.}\ \bibnamefont {Isakov}}, \bibinfo {author}
  {\bibfnamefont {E.}~\bibnamefont {Jeffrey}}, \bibinfo {author} {\bibfnamefont
  {Z.}~\bibnamefont {Jiang}}, \bibinfo {author} {\bibfnamefont
  {D.}~\bibnamefont {Kafri}}, \bibinfo {author} {\bibfnamefont
  {K.}~\bibnamefont {Kechedzhi}}, \bibinfo {author} {\bibfnamefont
  {T.}~\bibnamefont {Khattar}}, \bibinfo {author} {\bibfnamefont
  {S.}~\bibnamefont {Kim}}, \bibinfo {author} {\bibfnamefont {P.~V.}\
  \bibnamefont {Klimov}}, \bibinfo {author} {\bibfnamefont {A.~N.}\
  \bibnamefont {Korotkov}}, \bibinfo {author} {\bibfnamefont {F.}~\bibnamefont
  {Kostritsa}}, \bibinfo {author} {\bibfnamefont {D.}~\bibnamefont {Landhuis}},
  \bibinfo {author} {\bibfnamefont {P.}~\bibnamefont {Laptev}}, \bibinfo
  {author} {\bibfnamefont {A.}~\bibnamefont {Locharla}}, \bibinfo {author}
  {\bibfnamefont {E.}~\bibnamefont {Lucero}}, \bibinfo {author} {\bibfnamefont
  {O.}~\bibnamefont {Martin}}, \bibinfo {author} {\bibfnamefont {J.~R.}\
  \bibnamefont {McClean}}, \bibinfo {author} {\bibfnamefont {M.}~\bibnamefont
  {McEwen}}, \bibinfo {author} {\bibfnamefont {K.~C.}\ \bibnamefont {Miao}},
  \bibinfo {author} {\bibfnamefont {M.}~\bibnamefont {Mohseni}}, \bibinfo
  {author} {\bibfnamefont {S.}~\bibnamefont {Montazeri}}, \bibinfo {author}
  {\bibfnamefont {W.}~\bibnamefont {Mruczkiewicz}}, \bibinfo {author}
  {\bibfnamefont {J.}~\bibnamefont {Mutus}}, \bibinfo {author} {\bibfnamefont
  {O.}~\bibnamefont {Naaman}}, \bibinfo {author} {\bibfnamefont
  {M.}~\bibnamefont {Neeley}}, \bibinfo {author} {\bibfnamefont
  {C.}~\bibnamefont {Neill}}, \bibinfo {author} {\bibfnamefont {M.~Y.}\
  \bibnamefont {Niu}}, \bibinfo {author} {\bibfnamefont {T.~E.}\ \bibnamefont
  {O'Brien}}, \bibinfo {author} {\bibfnamefont {A.}~\bibnamefont {Opremcak}},
  \bibinfo {author} {\bibfnamefont {B.}~\bibnamefont {Pat{\'{o}}}}, \bibinfo
  {author} {\bibfnamefont {A.}~\bibnamefont {Petukhov}}, \bibinfo {author}
  {\bibfnamefont {N.~C.}\ \bibnamefont {Rubin}}, \bibinfo {author}
  {\bibfnamefont {D.}~\bibnamefont {Sank}}, \bibinfo {author} {\bibfnamefont
  {V.}~\bibnamefont {Shvarts}}, \bibinfo {author} {\bibfnamefont
  {D.}~\bibnamefont {Strain}}, \bibinfo {author} {\bibfnamefont
  {M.}~\bibnamefont {Szalay}}, \bibinfo {author} {\bibfnamefont
  {B.}~\bibnamefont {Villalonga}}, \bibinfo {author} {\bibfnamefont {T.~C.}\
  \bibnamefont {White}}, \bibinfo {author} {\bibfnamefont {Z.}~\bibnamefont
  {Yao}}, \bibinfo {author} {\bibfnamefont {P.}~\bibnamefont {Yeh}}, \bibinfo
  {author} {\bibfnamefont {J.}~\bibnamefont {Yoo}}, \bibinfo {author}
  {\bibfnamefont {A.}~\bibnamefont {Zalcman}}, \bibinfo {author} {\bibfnamefont
  {H.}~\bibnamefont {Neven}}, \bibinfo {author} {\bibfnamefont
  {S.}~\bibnamefont {Boixo}}, \bibinfo {author} {\bibfnamefont
  {A.}~\bibnamefont {Megrant}}, \bibinfo {author} {\bibfnamefont
  {Y.}~\bibnamefont {Chen}}, \bibinfo {author} {\bibfnamefont {J.}~\bibnamefont
  {Kelly}}, \bibinfo {author} {\bibfnamefont {V.}~\bibnamefont {Smelyanskiy}},
  \bibinfo {author} {\bibfnamefont {A.}~\bibnamefont {Kitaev}}, \bibinfo
  {author} {\bibfnamefont {M.}~\bibnamefont {Knap}}, \bibinfo {author}
  {\bibfnamefont {F.}~\bibnamefont {Pollmann}}, \ and\ \bibinfo {author}
  {\bibfnamefont {P.}~\bibnamefont {Roushan}},\ }\bibfield  {title} {\enquote
  {\bibinfo {title} {{Realizing topologically ordered states on a quantum
  processor}},}\ }\href {\doibase 10.1126/science.abi8378} {\bibfield
  {journal} {\bibinfo  {journal} {Science}\ }\textbf {\bibinfo {volume}
  {374}},\ \bibinfo {pages} {1237--1241} (\bibinfo {year} {2021})}\BibitemShut
  {NoStop}%
\end{thebibliography}%


\begin{thebibliography}{15}%
\makeatletter
\providecommand \@ifxundefined [1]{%
 \@ifx{#1\undefined}
}%
\providecommand \@ifnum [1]{%
 \ifnum #1\expandafter \@firstoftwo
 \else \expandafter \@secondoftwo
 \fi
}%
\providecommand \@ifx [1]{%
 \ifx #1\expandafter \@firstoftwo
 \else \expandafter \@secondoftwo
 \fi
}%
\providecommand \natexlab [1]{#1}%
\providecommand \enquote  [1]{``#1''}%
\providecommand \bibnamefont  [1]{#1}%
\providecommand \bibfnamefont [1]{#1}%
\providecommand \citenamefont [1]{#1}%
\providecommand \href@noop [0]{\@secondoftwo}%
\providecommand \href [0]{\begingroup \@sanitize@url \@href}%
\providecommand \@href[1]{\@@startlink{#1}\@@href}%
\providecommand \@@href[1]{\endgroup#1\@@endlink}%
\providecommand \@sanitize@url [0]{\catcode `\\12\catcode `\$12\catcode
  `\&12\catcode `\#12\catcode `\^12\catcode `\_12\catcode `\%12\relax}%
\providecommand \@@startlink[1]{}%
\providecommand \@@endlink[0]{}%
\providecommand \url  [0]{\begingroup\@sanitize@url \@url }%
\providecommand \@url [1]{\endgroup\@href {#1}{\urlprefix }}%
\providecommand \urlprefix  [0]{URL }%
\providecommand \Eprint [0]{\href }%
\providecommand \doibase [0]{http://dx.doi.org/}%
\providecommand \selectlanguage [0]{\@gobble}%
\providecommand \bibinfo  [0]{\@secondoftwo}%
\providecommand \bibfield  [0]{\@secondoftwo}%
\providecommand \translation [1]{[#1]}%
\providecommand \BibitemOpen [0]{}%
\providecommand \bibitemStop [0]{}%
\providecommand \bibitemNoStop [0]{.\EOS\space}%
\providecommand \EOS [0]{\spacefactor3000\relax}%
\providecommand \BibitemShut  [1]{\csname bibitem#1\endcsname}%
\let\auto@bib@innerbib\@empty
\bibitem [{\citenamefont {Monroe}\ \emph {et~al.}(2021)\citenamefont {Monroe},
  \citenamefont {Campbell}, \citenamefont {Duan}, \citenamefont {Gong},
  \citenamefont {Gorshkov}, \citenamefont {Hess}, \citenamefont {Islam},
  \citenamefont {Kim}, \citenamefont {Linke}, \citenamefont {Pagano},
  \citenamefont {Richerme}, \citenamefont {Senko},\ and\ \citenamefont
  {Yao}}]{Monroe2022}%
  \BibitemOpen
  \bibfield  {author} {\bibinfo {author} {\bibfnamefont {C.}~\bibnamefont
  {Monroe}}, \bibinfo {author} {\bibfnamefont {W.~C.}\ \bibnamefont
  {Campbell}}, \bibinfo {author} {\bibfnamefont {L.-M.}\ \bibnamefont {Duan}},
  \bibinfo {author} {\bibfnamefont {Z.-X.}\ \bibnamefont {Gong}}, \bibinfo
  {author} {\bibfnamefont {A.~V.}\ \bibnamefont {Gorshkov}}, \bibinfo {author}
  {\bibfnamefont {P.~W.}\ \bibnamefont {Hess}}, \bibinfo {author}
  {\bibfnamefont {R.}~\bibnamefont {Islam}}, \bibinfo {author} {\bibfnamefont
  {K.}~\bibnamefont {Kim}}, \bibinfo {author} {\bibfnamefont {N.~M.}\
  \bibnamefont {Linke}}, \bibinfo {author} {\bibfnamefont {G.}~\bibnamefont
  {Pagano}}, \bibinfo {author} {\bibfnamefont {P.}~\bibnamefont {Richerme}},
  \bibinfo {author} {\bibfnamefont {C.}~\bibnamefont {Senko}}, \ and\ \bibinfo
  {author} {\bibfnamefont {N.~Y.}\ \bibnamefont {Yao}},\ }\bibfield  {title}
  {\enquote {\bibinfo {title} {Programmable quantum simulations of spin systems
  with trapped ions},}\ }\href {\doibase 10.1103/RevModPhys.93.025001}
  {\bibfield  {journal} {\bibinfo  {journal} {Rev. Mod. Phys.}\ }\textbf
  {\bibinfo {volume} {93}},\ \bibinfo {pages} {025001} (\bibinfo {year}
  {2021})}\BibitemShut {NoStop}%
\bibitem [{\citenamefont {Feng}\ \emph {et~al.}(2022)\citenamefont {Feng},
  \citenamefont {Katz}, \citenamefont {Haack}, \citenamefont {Maghrebi},
  \citenamefont {Gorshkov}, \citenamefont {Gong}, \citenamefont {Cetina},\ and\
  \citenamefont {Monroe}}]{Lei2022}%
  \BibitemOpen
  \bibfield  {author} {\bibinfo {author} {\bibfnamefont {Lei}\ \bibnamefont
  {Feng}}, \bibinfo {author} {\bibfnamefont {Or}~\bibnamefont {Katz}}, \bibinfo
  {author} {\bibfnamefont {Casey}\ \bibnamefont {Haack}}, \bibinfo {author}
  {\bibfnamefont {Mohammad}\ \bibnamefont {Maghrebi}}, \bibinfo {author}
  {\bibfnamefont {Alexey~V.}\ \bibnamefont {Gorshkov}}, \bibinfo {author}
  {\bibfnamefont {Zhexuan}\ \bibnamefont {Gong}}, \bibinfo {author}
  {\bibfnamefont {Marko}\ \bibnamefont {Cetina}}, \ and\ \bibinfo {author}
  {\bibfnamefont {Christopher}\ \bibnamefont {Monroe}},\ }\bibfield  {title}
  {\enquote {\bibinfo {title} {{Continuous Symmetry Breaking in a Trapped-Ion
  Spin Chain}},}\ }\href {https://arxiv.org/abs/2211.01275} {\bibfield
  {journal} {\bibinfo  {journal} {arXiv:2211.01275}\ } (\bibinfo {year}
  {2022})}\BibitemShut {NoStop}%
\bibitem [{\citenamefont {Katz}\ \emph {et~al.}(2023)\citenamefont {Katz},
  \citenamefont {Feng}, \citenamefont {Risinger}, \citenamefont {Monroe},\ and\
  \citenamefont {Cetina}}]{katz2023}%
  \BibitemOpen
  \bibfield  {author} {\bibinfo {author} {\bibfnamefont {Or}~\bibnamefont
  {Katz}}, \bibinfo {author} {\bibfnamefont {Lei}\ \bibnamefont {Feng}},
  \bibinfo {author} {\bibfnamefont {Andrew}\ \bibnamefont {Risinger}}, \bibinfo
  {author} {\bibfnamefont {Christopher}\ \bibnamefont {Monroe}}, \ and\
  \bibinfo {author} {\bibfnamefont {Marko}\ \bibnamefont {Cetina}},\ }\bibfield
   {title} {\enquote {\bibinfo {title} {Demonstration of three- and four-body
  interactions between trapped-ion spins},}\ }\href {\doibase
  10.1038/s41567-023-02102-7} {\bibfield  {journal} {\bibinfo  {journal}
  {Nature Physics}\ }\textbf {\bibinfo {volume} {19}},\ \bibinfo {pages}
  {1452--1458} (\bibinfo {year} {2023})}\BibitemShut {NoStop}%
\bibitem [{\citenamefont {Morong}\ \emph {et~al.}(2023)\citenamefont {Morong},
  \citenamefont {Collins}, \citenamefont {De}, \citenamefont {Stavropoulos},
  \citenamefont {You},\ and\ \citenamefont {Monroe}}]{morong2023engineering}%
  \BibitemOpen
  \bibfield  {author} {\bibinfo {author} {\bibfnamefont {W.}~\bibnamefont
  {Morong}}, \bibinfo {author} {\bibfnamefont {K.S.}\ \bibnamefont {Collins}},
  \bibinfo {author} {\bibfnamefont {A.}~\bibnamefont {De}}, \bibinfo {author}
  {\bibfnamefont {E.}~\bibnamefont {Stavropoulos}}, \bibinfo {author}
  {\bibfnamefont {T.}~\bibnamefont {You}}, \ and\ \bibinfo {author}
  {\bibfnamefont {C.}~\bibnamefont {Monroe}},\ }\bibfield  {title} {\enquote
  {\bibinfo {title} {{Engineering Dynamically Decoupled Quantum Simulations
  with Trapped Ions}},}\ }\href {\doibase 10.1103/PRXQuantum.4.010334}
  {\bibfield  {journal} {\bibinfo  {journal} {PRX Quantum}\ }\textbf {\bibinfo
  {volume} {4}},\ \bibinfo {pages} {010334} (\bibinfo {year}
  {2023})}\BibitemShut {NoStop}%
\bibitem [{\citenamefont {Barthel}(2016)}]{PhysRevB.94.115157}%
  \BibitemOpen
  \bibfield  {author} {\bibinfo {author} {\bibfnamefont {Thomas}\ \bibnamefont
  {Barthel}},\ }\bibfield  {title} {\enquote {\bibinfo {title} {{Matrix product
  purifications for canonical ensembles and quantum number distributions}},}\
  }\href {\doibase 10.1103/PhysRevB.94.115157} {\bibfield  {journal} {\bibinfo
  {journal} {Phys. Rev. B}\ }\textbf {\bibinfo {volume} {94}},\ \bibinfo
  {pages} {115157} (\bibinfo {year} {2016})}\BibitemShut {NoStop}%
\bibitem [{\citenamefont {Hauschild}\ \emph {et~al.}(2018)\citenamefont
  {Hauschild}, \citenamefont {Leviatan}, \citenamefont {Bardarson},
  \citenamefont {Altman}, \citenamefont {Zaletel},\ and\ \citenamefont
  {Pollmann}}]{PhysRevB.98.235163}%
  \BibitemOpen
  \bibfield  {author} {\bibinfo {author} {\bibfnamefont {Johannes}\
  \bibnamefont {Hauschild}}, \bibinfo {author} {\bibfnamefont {Eyal}\
  \bibnamefont {Leviatan}}, \bibinfo {author} {\bibfnamefont {Jens~H.}\
  \bibnamefont {Bardarson}}, \bibinfo {author} {\bibfnamefont {Ehud}\
  \bibnamefont {Altman}}, \bibinfo {author} {\bibfnamefont {Michael~P.}\
  \bibnamefont {Zaletel}}, \ and\ \bibinfo {author} {\bibfnamefont {Frank}\
  \bibnamefont {Pollmann}},\ }\bibfield  {title} {\enquote {\bibinfo {title}
  {Finding purifications with minimal entanglement},}\ }\href {\doibase
  10.1103/PhysRevB.98.235163} {\bibfield  {journal} {\bibinfo  {journal} {Phys.
  Rev. B}\ }\textbf {\bibinfo {volume} {98}},\ \bibinfo {pages} {235163}
  (\bibinfo {year} {2018})}\BibitemShut {NoStop}%
\bibitem [{\citenamefont {Zaletel}\ \emph {et~al.}(2015)\citenamefont
  {Zaletel}, \citenamefont {Mong}, \citenamefont {Karrasch}, \citenamefont
  {Moore},\ and\ \citenamefont {Pollmann}}]{PhysRevB.91.165112}%
  \BibitemOpen
  \bibfield  {author} {\bibinfo {author} {\bibfnamefont {Michael~P.}\
  \bibnamefont {Zaletel}}, \bibinfo {author} {\bibfnamefont {Roger S.~K.}\
  \bibnamefont {Mong}}, \bibinfo {author} {\bibfnamefont {Christoph}\
  \bibnamefont {Karrasch}}, \bibinfo {author} {\bibfnamefont {Joel~E.}\
  \bibnamefont {Moore}}, \ and\ \bibinfo {author} {\bibfnamefont {Frank}\
  \bibnamefont {Pollmann}},\ }\bibfield  {title} {\enquote {\bibinfo {title}
  {Time-evolving a matrix product state with long-ranged interactions},}\
  }\href {\doibase 10.1103/PhysRevB.91.165112} {\bibfield  {journal} {\bibinfo
  {journal} {Phys. Rev. B}\ }\textbf {\bibinfo {volume} {91}},\ \bibinfo
  {pages} {165112} (\bibinfo {year} {2015})}\BibitemShut {NoStop}%
\bibitem [{\citenamefont {Hauschild}\ and\ \citenamefont
  {Pollmann}(2018)}]{tenpy}%
  \BibitemOpen
  \bibfield  {author} {\bibinfo {author} {\bibfnamefont {Johannes}\
  \bibnamefont {Hauschild}}\ and\ \bibinfo {author} {\bibfnamefont {Frank}\
  \bibnamefont {Pollmann}},\ }\bibfield  {title} {\enquote {\bibinfo {title}
  {{Efficient numerical simulations with Tensor Networks: Tensor Network Python
  (TeNPy)}},}\ }\href {\doibase 10.21468/SciPostPhysLectNotes.5} {\bibfield
  {journal} {\bibinfo  {journal} {SciPost Phys. Lect. Notes}\ ,\ \bibinfo
  {pages} {5}} (\bibinfo {year} {2018})},\ \bibinfo {note} {code available from
  \url{https://github.com/tenpy/tenpy}}\BibitemShut {NoStop}%
\bibitem [{\citenamefont {Wolff}(1989)}]{PhysRevLett.62.361}%
  \BibitemOpen
  \bibfield  {author} {\bibinfo {author} {\bibfnamefont {Ulli}\ \bibnamefont
  {Wolff}},\ }\bibfield  {title} {\enquote {\bibinfo {title} {{Collective Monte
  Carlo Updating for Spin Systems}},}\ }\href {\doibase
  10.1103/PhysRevLett.62.361} {\bibfield  {journal} {\bibinfo  {journal} {Phys.
  Rev. Lett.}\ }\textbf {\bibinfo {volume} {62}},\ \bibinfo {pages} {361--364}
  (\bibinfo {year} {1989})}\BibitemShut {NoStop}%
\bibitem [{\citenamefont {Luijten}\ and\ \citenamefont
  {Bl\"{o}te}(1995)}]{LUIJTEN1995}%
  \BibitemOpen
  \bibfield  {author} {\bibinfo {author} {\bibfnamefont {Erik}\ \bibnamefont
  {Luijten}}\ and\ \bibinfo {author} {\bibfnamefont {Henk~W.J.}\ \bibnamefont
  {Bl\"{o}te}},\ }\bibfield  {title} {\enquote {\bibinfo {title} {{Monte Carlo
  method for spin models with long-range interactions}},}\ }\href {\doibase
  10.1142/s0129183195000265} {\bibfield  {journal} {\bibinfo  {journal}
  {International Journal of Modern Physics C}\ }\textbf {\bibinfo {volume}
  {06}},\ \bibinfo {pages} {359--370} (\bibinfo {year} {1995})}\BibitemShut
  {NoStop}%
\bibitem [{\citenamefont {Lazo}\ \emph {et~al.}(2021)\citenamefont {Lazo},
  \citenamefont {Heyl}, \citenamefont {Dalmonte},\ and\ \citenamefont
  {Angelone}}]{GonzalezLazo2021}%
  \BibitemOpen
  \bibfield  {author} {\bibinfo {author} {\bibfnamefont {Eduardo~Gonzalez}\
  \bibnamefont {Lazo}}, \bibinfo {author} {\bibfnamefont {Markus}\ \bibnamefont
  {Heyl}}, \bibinfo {author} {\bibfnamefont {Marcello}\ \bibnamefont
  {Dalmonte}}, \ and\ \bibinfo {author} {\bibfnamefont {Adriano}\ \bibnamefont
  {Angelone}},\ }\bibfield  {title} {\enquote {\bibinfo {title}
  {{Finite-temperature critical behavior of long-range quantum Ising
  models}},}\ }\href {\doibase 10.21468/SciPostPhys.11.4.076} {\bibfield
  {journal} {\bibinfo  {journal} {SciPost Phys.}\ }\textbf {\bibinfo {volume}
  {11}},\ \bibinfo {pages} {076} (\bibinfo {year} {2021})}\BibitemShut
  {NoStop}%
\bibitem [{\citenamefont {Desai}\ \emph {et~al.}(1988)\citenamefont {Desai},
  \citenamefont {Heermann},\ and\ \citenamefont {Binder}}]{Desai1988}%
  \BibitemOpen
  \bibfield  {author} {\bibinfo {author} {\bibfnamefont {Rashmi~C.}\
  \bibnamefont {Desai}}, \bibinfo {author} {\bibfnamefont {Dieter~W.}\
  \bibnamefont {Heermann}}, \ and\ \bibinfo {author} {\bibfnamefont
  {K.}~\bibnamefont {Binder}},\ }\bibfield  {title} {\enquote {\bibinfo {title}
  {Finite-size scaling in a microcanonical ensemble},}\ }\href {\doibase
  10.1007/bf01014226} {\bibfield  {journal} {\bibinfo  {journal} {Journal of
  Statistical Physics}\ }\textbf {\bibinfo {volume} {53}},\ \bibinfo {pages}
  {795--823} (\bibinfo {year} {1988})}\BibitemShut {NoStop}%
\bibitem [{\citenamefont {Korenblit}\ \emph {et~al.}(2012)\citenamefont
  {Korenblit}, \citenamefont {Kafri}, \citenamefont {Campbell}, \citenamefont
  {Islam}, \citenamefont {Edwards}, \citenamefont {Gong}, \citenamefont {Lin},
  \citenamefont {Duan}, \citenamefont {Kim}, \citenamefont {Kim},\ and\
  \citenamefont {Monroe}}]{Korenblit2012}%
  \BibitemOpen
  \bibfield  {author} {\bibinfo {author} {\bibfnamefont {S}~\bibnamefont
  {Korenblit}}, \bibinfo {author} {\bibfnamefont {D}~\bibnamefont {Kafri}},
  \bibinfo {author} {\bibfnamefont {W~C}\ \bibnamefont {Campbell}}, \bibinfo
  {author} {\bibfnamefont {R}~\bibnamefont {Islam}}, \bibinfo {author}
  {\bibfnamefont {E~E}\ \bibnamefont {Edwards}}, \bibinfo {author}
  {\bibfnamefont {Z-X}\ \bibnamefont {Gong}}, \bibinfo {author} {\bibfnamefont
  {G-D}\ \bibnamefont {Lin}}, \bibinfo {author} {\bibfnamefont {L-M}\
  \bibnamefont {Duan}}, \bibinfo {author} {\bibfnamefont {J}~\bibnamefont
  {Kim}}, \bibinfo {author} {\bibfnamefont {K}~\bibnamefont {Kim}}, \ and\
  \bibinfo {author} {\bibfnamefont {C}~\bibnamefont {Monroe}},\ }\bibfield
  {title} {\enquote {\bibinfo {title} {Quantum simulation of spin models on an
  arbitrary lattice with trapped ions},}\ }\href {\doibase
  10.1088/1367-2630/14/9/095024} {\bibfield  {journal} {\bibinfo  {journal}
  {New Journal of Physics}\ }\textbf {\bibinfo {volume} {14}},\ \bibinfo
  {pages} {095024} (\bibinfo {year} {2012})}\BibitemShut {NoStop}%
\bibitem [{\citenamefont {Srednicki}(1996)}]{Srednicki1996}%
  \BibitemOpen
  \bibfield  {author} {\bibinfo {author} {\bibfnamefont {Mark}\ \bibnamefont
  {Srednicki}},\ }\bibfield  {title} {\enquote {\bibinfo {title} {Thermal
  fluctuations in quantized chaotic systems},}\ }\href {\doibase
  10.1088/0305-4470/29/4/003} {\bibfield  {journal} {\bibinfo  {journal}
  {Journal of Physics A: Mathematical and General}\ }\textbf {\bibinfo {volume}
  {29}},\ \bibinfo {pages} {L75--L79} (\bibinfo {year} {1996})}\BibitemShut
  {NoStop}%
\bibitem [{\citenamefont {Zhang}\ \emph {et~al.}(2017)\citenamefont {Zhang},
  \citenamefont {Pagano}, \citenamefont {Hess}, \citenamefont {Kyprianidis},
  \citenamefont {Becker}, \citenamefont {Kaplan}, \citenamefont {Gorshkov},
  \citenamefont {Gong},\ and\ \citenamefont {Monroe}}]{Zhang20172}%
  \BibitemOpen
  \bibfield  {author} {\bibinfo {author} {\bibfnamefont {J.}~\bibnamefont
  {Zhang}}, \bibinfo {author} {\bibfnamefont {G.}~\bibnamefont {Pagano}},
  \bibinfo {author} {\bibfnamefont {P.~W.}\ \bibnamefont {Hess}}, \bibinfo
  {author} {\bibfnamefont {A.}~\bibnamefont {Kyprianidis}}, \bibinfo {author}
  {\bibfnamefont {P.}~\bibnamefont {Becker}}, \bibinfo {author} {\bibfnamefont
  {H.}~\bibnamefont {Kaplan}}, \bibinfo {author} {\bibfnamefont {A.~V.}\
  \bibnamefont {Gorshkov}}, \bibinfo {author} {\bibfnamefont {Z.-X.}\
  \bibnamefont {Gong}}, \ and\ \bibinfo {author} {\bibfnamefont
  {C.}~\bibnamefont {Monroe}},\ }\bibfield  {title} {\enquote {\bibinfo {title}
  {{Observation of a many-body dynamical phase transition with a 53-qubit
  quantum simulator}},}\ }\href {\doibase 10.1038/nature24654} {\bibfield
  {journal} {\bibinfo  {journal} {Nature}\ }\textbf {\bibinfo {volume} {551}},\
  \bibinfo {pages} {601--604} (\bibinfo {year} {2017})}\BibitemShut {NoStop}%
\end{thebibliography}%

\end{document}


\title{Supplementary materials for: Observation of a finite-energy phase transition in a one-dimensional quantum simulator}
\author{Alexander Schuckert$^{*\dagger}$}
\affiliation{Joint Quantum Institute and Joint Center for Quantum Information and Computer Science, University of Maryland and NIST, College Park, Maryland 20742, USA}
\author{Or Katz$^{*\dagger}$}
\affiliation{Duke Quantum Center, Department of Physics and Electrical and Computer Engineering, Duke University, Durham, NC 27701}
\author{Lei Feng}
\affiliation{Duke Quantum Center, Department of Physics and Electrical and Computer Engineering, Duke University, Durham, NC 27701}
\author{Eleanor Crane}
\affiliation{Joint Quantum Institute and Joint Center for Quantum Information and Computer Science, University of Maryland and NIST, College Park, Maryland 20742, USA}
\author{Arinjoy De}
\affiliation{Joint Quantum Institute and Joint Center for Quantum Information and Computer Science, University of Maryland and NIST, College Park, Maryland 20742, USA}
\author{Mohammad Hafezi}
\affiliation{Joint Quantum Institute and Joint Center for Quantum Information and Computer Science, University of Maryland and NIST, College Park, Maryland 20742, USA}
\author{Alexey V.~Gorshkov}
\affiliation{Joint Quantum Institute and Joint Center for Quantum Information and Computer Science, University of Maryland and NIST, College Park, Maryland 20742, USA}
\author{Christopher Monroe}
\affiliation{Duke Quantum Center, Department of Physics and Electrical and Computer Engineering, Duke University, Durham, NC 27701}

\maketitle
\tableofcontents
 \beginsupplement
\section{Experimental details}

In this section, we introduce the trapped ion experiment in more detail, specifically emphasizing how we prepare the interactions in Eq.~(1) of the main text and how we choose the initial states.

\subsection{Generating spin-spin interactions between ions}

We induce spin-spin interactions among trapped ions through Raman transitions that virtually excite the collective motion of the ions. These Raman transitions are generated using pairs of beams: one that globally addresses the ion chain and another that individually targets each ion. The global addressing beam passes through an acousto-optical modulator (AOM) concurrently driven by two radio-frequency (RF) signals. This process splits the optical beam into two components, each with a distinct tone and nearly equal power, which are then projected onto the ion crystal. Simultaneously, a perpendicular array of tightly focused beams is directed toward the ion positions. Precise control of the trapping potential ensures a high degree of overlap between each ion and these beams. We achieve simultaneous and independent control over the amplitudes and frequencies of these beams using a multi-channel AOM. This configuration drives both the first red- and blue-sideband transitions within the dispersive regime. The beatnote frequencies for these transitions are detuned by $\pm(\omega_{N}+\Delta_{\pm})$ from the carrier transition, where $\omega_{N}$ is the lowest frequency of the phonon mode along the radial direction. The detunings are nearly symmetric $|\Delta_{+}-\Delta_{-}|\ll|\Delta_{\pm}|$.

\begin{figure}
    \centering
    \includegraphics{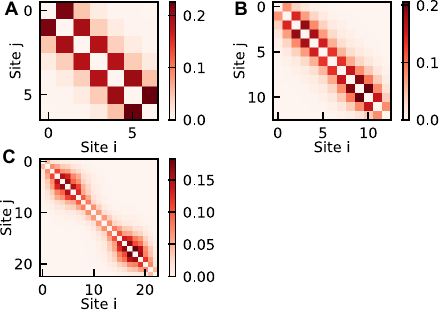}
    \caption{\textbf{Interaction matrices.} Normalised experimentally realized interaction matrices $J_{ij}/\mathcal{N}$ calculated from the mode spectrum and the beam parameters. (\textbf{A}) L=7, (\textbf{B}) L=13, (\textbf{C}) L=23.}
    \label{fig:expinteractions}
\end{figure}

This configuration generates the Hamiltonian in Eq.~(1) for a chain of $N$ ions. The asymmetry in the detunings between the two tones leads to an effective transverse magnetic field with an amplitude $g=(\Delta_{+}-\Delta_{-})/4$ in the frame that rotates at the carrier frequency. The symmetric part, $\Delta = (\Delta_{+}-\Delta_{-})/2$, corresponds to the average detuning of the effective spin-dependent force. This, in turn, generates the Ising Hamiltonian with an interaction matrix \cite{Monroe2022,Lei2022}
\begin{equation} \label{eq:Jij_formula}J_{ij}=\sum_{k}\frac{\eta_{ik}\eta_{jk}\Omega_{i}\Omega_{j}}{2(\Delta-\omega_{N}-\omega_{k})}.
\end{equation}
Here, $\eta_{ik}=0.08b_{ik}$ are the Lamb-Dicke parameters, with $b_{ik}$ as the mode participation matrix elements describing the coupling between spin $i$ and motional mode $k$ \cite{katz2023}. $\Omega_{i}$ denotes the equivalent resonant carrier Rabi frequency at ion $1\leq i \leq N$ for each tone, and $\omega_{k}$ represents the motional frequencies along one radial direction (labeled in decreasing order with $1\leq k \leq N$). These frequencies are determined by the trapping potential; we employ a quadratic trapping potential in the radial direction with center-of-mass frequency of $\omega_{1}=2\pi\times3.075$ MHz and an axial electrostatic potential of $V(x)=c_4 x^{4}+c_2 x^{2}$, where $c_2=0.11\, \textrm{eV/mm}^2, c_4=1.6\times10^3\,\textrm{eV/mm}^4$ for a $15$-ion chain, and $c_2=-0.1\,\textrm{eV/mm}^2,  c_4=235\,\textrm{eV/mm}^4$ for a $27$-ion chain. Here, $x$ is the coordinate along the chain axis.  These potentials result in a nearly uniform spacing of the central ions in the crystal, except for the edges, with a spacing of $3.75\,\mu$m to match the centers of the uniformly spaced beams used for individual addressing.
The effective wave-vector of the optical field is aligned to selectively drive only one specific set of radial phonon modes. 

To estimate the form of the $J_{ij}$, we calculate the mode participation factors by solving the Laplace equation assuming harmonic radial potential with a strength measured from experiment and used the electrode voltages, from which we find the positions of the ions, from which we determine the mode participation matrix elements and the eigenfrequencies. This then enables us to calculate the $J_{ij}$ from the measured individual Rabi frequencies and the detunings from the modes using Eq.~\eqref{eq:Jij_formula}, see Ref.~\cite{Lei2022} for more details on our setup.

We detune $\Delta$ to the red side of the mode spectrum ($\Delta<0$) to primarily couple to phonon
modes characterized by rapidly varying mode participation factors (i.e., near the zig-zag phonon mode and far from the center-of-mass phonon mode). For a uniform-spaced ion chain, this choice results in an interaction matrix $J_{ij}$ consisting of two terms primarily dependent on the ion separation distance  $|i-j|$: an inverse cubic (power-law) term and an exponentially decreasing term that alternates with the spin distance, appended by a factor $(-1)^{|i-j|}$. As we reduce $\Delta$, the exponential term becomes dominant over the power-law term, particularly at short distances. To generate a non-alternating $J_{ij}$ matrix, we shift the optical phase of the individual beam array in a staggered way, shifting the phase of all odd beams by $\pi$.

\begin{figure}
     \centering
     \includegraphics[width=\columnwidth]{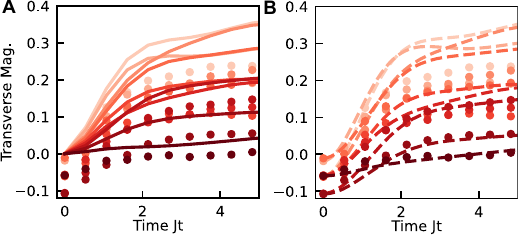}
     \caption{\textbf{Dynamics of the transverse magnetization $\sum_i\braket{\hat \sigma^z_i}/L$.} (\textbf{A}) Numerical simulations using the same $J_{ij}$ as the experiment (solid lines). (\textbf{B}) Numerical simulations using a tilted initial state (dashed lines, see text for definition). Dots in both subfigures are experimental data for the same initial states as shown in Fig.~2 of the main text. $g/J=0.31$. $L=13$.}
     \label{fig:transverse}
 \end{figure}

Individual control over the beam amplitudes provides additional flexibility in manipulating the interaction matrix. Specifically, by turning off the beam targeting ion $q$, we effectively eliminate $J_{qi}$ for all $1\leq i \leq N$, rendering its participation in the phonon modes independent of its spin. This capability allows us to simulate the evolution of $L\leq N$ spins in an $N$-ion chain. Utilizing this technique, we realize the interaction of $L=7,13,23$ spins in chains of $N=15,15,27$ ions, respectively, with the beams symmetrically turned off near the crystal's edges. Not including the edge ions' spins allows us to achieve chains with more uniform spacing through the presence of their charge. 

For the $L=7,13$ configurations, we employ uniform beam amplitudes. In the case of the $L=23$ configuration, the relative Rabi frequencies of the $23$ beams are chosen as follows: $ [1,1,0.72,0.76,0.6,0.72,0.63,0.81,0.72,0.91,0.78,0.94,$ $0.78 ,0.9,0.72,0.8,0.62,0.71,0.59,0.75,0.71,0.99,1]$, alleviating the variation of the interaction strength across the chain due to the spatial dependence of the rapidly varying mode participation factors. We define $J=\textrm{max}_{ij}(J_{ij})$ and had $J\approx 1 \times 2\pi$ kHz in all experiments. The detunings are set at $\Delta=2\pi \times 100$ kHz for $L=7$, $\Delta=2\pi \times  35$ kHz for $L=13$, and $\Delta=2\pi \times  9$ kHz for $L=23$.

To account for the light shift induced by the beams on the ions in each configuration, we perform calibration and compensation by applying an opposite shift to the optical frequencies of the individual beams. For the $L=13$ configuration, we additionally employ a variation of a dynamical decoupling technique \cite{morong2023engineering} that we have found effective in mitigating $\sigma_z$ noise (such as light-shift noise). In particular, we break the Hamiltonian evolution into two periods, separated by a short $\pi$ pulse that is applied to all spins, rotating around the $x$-axis. Inverting the sign of $g$ for the second evolution period yields the same evolution as the original Hamiltonian.

\subsection{Initial-state preparation\label{sct:stateprep}}

The initial states are chosen by selecting target energies $E_\mathrm{target}$ equally spaced in a window $[E_\mathrm{min},E_\mathrm{max}]$, where $E_\mathrm{min}$ is the energy of the totally polarized state (i.e.~the ground state for $g=0$) and  $E_\mathrm{max}$ is some high-energy limit close to zero. We then find the x-product state $\ket{\psi}$ which minimizes $|\braket{\psi|\hat H|\psi}-E_\mathrm{target}|$ for all $E_\mathrm{target}$, using the experimental $J_{ij}$. In practice, we do this by calculating $\braket{\psi|\hat H|\psi}$ for all $2^L$ states $\ket{\psi}$, which is numerically cheap even for $L=23$. For much larger systems, finding states closest to a particular energy could be done by using an optimisation algorithm. We show the thus selected initial states in tables~\ref{tab:states7},~\ref{tab:states13},~\ref{tab:states23}. Note that states with the same number of spin flips can have different energies due to the lack of translational invariance and the inhomogeneities present in the experimentally realized $J_{ij}$, see Fig.~\ref{fig:expinteractions}.

\begin{table}[]
    \centering
    \begin{tabular}{c|c}
    Energy density $\epsilon/J$ & State\\
-0.43&$\downarrow \downarrow \downarrow \downarrow \downarrow \downarrow \downarrow$\\
       -0.26&$\downarrow \downarrow \downarrow \downarrow \downarrow \downarrow \uparrow$\\
       -0.25&$\downarrow \downarrow \downarrow \downarrow \downarrow \uparrow \downarrow$\\
       -0.24&$\downarrow \downarrow \downarrow \downarrow \downarrow \uparrow \uparrow$\\
       -0.152&$\downarrow \downarrow \downarrow \downarrow \uparrow \downarrow \downarrow$\\
      -0.151& $\downarrow \downarrow \downarrow \downarrow \uparrow \downarrow \uparrow$\\
       -0.146&$\downarrow \downarrow \downarrow \downarrow \uparrow \uparrow \downarrow$\\
       -0.1&$\downarrow \downarrow \downarrow \downarrow \uparrow \uparrow \uparrow$\\
       -0.04&$\downarrow \downarrow \downarrow \uparrow \downarrow \downarrow \downarrow$\\
       0.0008&$\downarrow \downarrow \downarrow \uparrow \downarrow \downarrow \uparrow$
    \end{tabular}
    \caption{\textbf{Initial states for $L=7$.} States are eigenstates of $\hat \sigma^x_i$.}
    \label{tab:states7}
\end{table}

\begin{table}[]
    \centering
    \begin{tabular}{c|c}
    Energy density $\epsilon/J$ & State\\
-0.46&$\downarrow \downarrow \downarrow \downarrow \downarrow \downarrow \downarrow \downarrow \downarrow \downarrow \downarrow \downarrow \downarrow$\\
       -0.41&$\downarrow \downarrow \downarrow \downarrow \downarrow \downarrow \downarrow \downarrow \downarrow \downarrow \downarrow \downarrow \uparrow$\\
       -0.31&$\downarrow \downarrow \downarrow \downarrow \downarrow \downarrow \uparrow \downarrow \downarrow \downarrow \downarrow \downarrow \downarrow$\\
       -0.27&$\downarrow \downarrow \downarrow \downarrow \downarrow \downarrow \downarrow \downarrow \downarrow \uparrow \downarrow \downarrow \downarrow$\\
       -0.22&$\downarrow \uparrow \uparrow \uparrow \uparrow \uparrow \downarrow \uparrow \uparrow \uparrow \uparrow \uparrow \downarrow$\\
       -0.19&$\downarrow \downarrow \downarrow \uparrow \uparrow \downarrow \uparrow \uparrow \uparrow \uparrow \uparrow \uparrow \uparrow$\\
       -0.13&$\downarrow \downarrow \downarrow \uparrow \uparrow \uparrow \uparrow \uparrow \downarrow \uparrow \uparrow \uparrow \uparrow$\\
       -0.09&$\downarrow \downarrow \uparrow \downarrow \uparrow \uparrow \uparrow \uparrow \downarrow \downarrow \downarrow \downarrow \downarrow$\\
       -0.04&$\downarrow \uparrow \downarrow \downarrow \uparrow \uparrow \downarrow \downarrow \uparrow \uparrow \uparrow \uparrow \uparrow$
    \end{tabular}
    \caption{\textbf{Initial states for $L=13$. States are eigenstates of $\hat \sigma^x_i$.}}
    \label{tab:states13}
\end{table}
\begin{table}[]
    \centering
    \begin{tabular}{c|c}
    Energy density $\epsilon/J$ & State\\
-0.48&$\downarrow \downarrow \downarrow \downarrow \downarrow \downarrow \downarrow \downarrow \downarrow \downarrow \downarrow \downarrow \downarrow \downarrow \downarrow \downarrow \downarrow \downarrow \downarrow \downarrow \downarrow \downarrow
        \downarrow$\\
        -0.36&$\downarrow \downarrow \downarrow \uparrow \downarrow \downarrow \downarrow \downarrow \downarrow \downarrow \downarrow \downarrow \downarrow \downarrow \downarrow \downarrow \downarrow \downarrow \downarrow \downarrow \downarrow \downarrow
        \downarrow$\\
        -0.28&$\downarrow \downarrow \downarrow \downarrow \downarrow \downarrow \downarrow \downarrow \downarrow \downarrow \downarrow \downarrow \downarrow \downarrow \downarrow \downarrow \downarrow \downarrow \uparrow \uparrow \downarrow \downarrow
        \downarrow$\\
        -0.25&$\downarrow \downarrow \downarrow \downarrow \downarrow \uparrow \uparrow \downarrow \uparrow \downarrow \downarrow \downarrow \downarrow \downarrow \downarrow \downarrow \downarrow \downarrow \downarrow \downarrow \downarrow \downarrow
        \downarrow$\\
        -0.20&$\downarrow \downarrow \downarrow \downarrow \uparrow \uparrow \downarrow \downarrow \uparrow \uparrow \downarrow \uparrow \downarrow \downarrow \downarrow \downarrow \downarrow \downarrow \downarrow \downarrow \downarrow \downarrow
        \downarrow$\\
        -0.18&$\downarrow \downarrow \downarrow \uparrow \downarrow \downarrow \uparrow \uparrow \downarrow \downarrow \downarrow \uparrow \uparrow \downarrow \downarrow \downarrow \downarrow \downarrow \downarrow \downarrow \downarrow \downarrow
        \downarrow$\\
        -0.169&$\downarrow \downarrow \downarrow \uparrow \downarrow \downarrow \downarrow \uparrow \uparrow \downarrow \downarrow \downarrow \uparrow \uparrow \downarrow \downarrow \downarrow \downarrow \downarrow \downarrow \downarrow \downarrow
        \downarrow$\\
       -0.167&$\downarrow \downarrow \downarrow \uparrow \uparrow \downarrow \downarrow \uparrow \uparrow \downarrow \downarrow \uparrow \uparrow \downarrow \downarrow \downarrow \downarrow \downarrow \downarrow \downarrow \downarrow \downarrow
        \downarrow$\\
        -0.08&$\downarrow \downarrow \downarrow \downarrow \uparrow \uparrow \downarrow \downarrow \uparrow \uparrow \downarrow \downarrow \uparrow \downarrow \downarrow \uparrow \downarrow \downarrow \downarrow \downarrow \downarrow \downarrow
        \downarrow$\\
        -0.05&$\downarrow \downarrow \downarrow \uparrow \downarrow \downarrow \downarrow \downarrow \downarrow \downarrow \downarrow \downarrow \uparrow \uparrow \downarrow \uparrow \downarrow \downarrow \uparrow \uparrow \downarrow \uparrow
        \downarrow$\\
        -0.03&$\downarrow \downarrow \downarrow \downarrow \uparrow \downarrow \downarrow \uparrow \uparrow \downarrow \downarrow \uparrow \downarrow \uparrow \uparrow \downarrow \downarrow \downarrow \downarrow \downarrow \uparrow \uparrow
        \uparrow$\\
       -0.02&$\downarrow \downarrow \downarrow \uparrow \downarrow \downarrow \uparrow \downarrow \downarrow \uparrow \downarrow \uparrow \uparrow \uparrow \uparrow \uparrow \uparrow \uparrow \downarrow \downarrow \uparrow \uparrow
        \downarrow$\\
        0.001&$\downarrow \downarrow \downarrow \downarrow \downarrow \downarrow \uparrow \uparrow \downarrow \downarrow \downarrow \uparrow \uparrow \uparrow \downarrow \uparrow \downarrow \downarrow \uparrow \uparrow \downarrow \downarrow
        \uparrow$\\
        0.02&$\downarrow \downarrow \downarrow \uparrow \downarrow \downarrow \downarrow \downarrow \uparrow \uparrow \downarrow \uparrow \uparrow \uparrow \downarrow \downarrow \uparrow \uparrow \downarrow \downarrow \uparrow \uparrow
        \downarrow$\\
        0.04&$\downarrow \downarrow \downarrow \downarrow \uparrow \downarrow \downarrow \uparrow \uparrow \downarrow \downarrow \uparrow \downarrow \downarrow \uparrow \downarrow \downarrow \uparrow \downarrow \uparrow \downarrow \downarrow
        \downarrow$\\
        0.05&$\downarrow \downarrow \downarrow \downarrow \uparrow \uparrow \downarrow \downarrow \uparrow \uparrow \uparrow \uparrow \downarrow \uparrow \downarrow \uparrow \uparrow \downarrow \downarrow \uparrow \uparrow \downarrow
        \uparrow$\\
       0.0679&$\downarrow \downarrow \downarrow \uparrow \uparrow \downarrow \downarrow \uparrow \uparrow \downarrow \downarrow \uparrow \uparrow \downarrow \downarrow \downarrow \uparrow \uparrow \downarrow \uparrow \uparrow \downarrow
        \uparrow$\\
        0.0683&$\downarrow \downarrow \downarrow \downarrow \uparrow \uparrow \downarrow \downarrow \uparrow \uparrow \downarrow \uparrow \downarrow \downarrow \uparrow \downarrow \downarrow \uparrow \uparrow \downarrow \downarrow \uparrow
        \uparrow$\\
        0.074&$\downarrow \downarrow \downarrow \uparrow \uparrow \downarrow \downarrow \uparrow \uparrow \downarrow \downarrow \uparrow \uparrow \downarrow \downarrow \downarrow \uparrow \uparrow \downarrow \downarrow \uparrow \uparrow
        \downarrow$
    \end{tabular}
    \caption{\textbf{Initial states for $L=23$.} States are eigenstates of $\hat \sigma^x_i$.}
    \label{tab:states23}
\end{table}

As visible in Fig.~3 of the main text, we found less good agreement between theory and experiment for the transverse magnetization than for the squared magnetization. We attribute at least some of this additional error to over-underrotation errors in the state preparation. Indeed, we observed that the initial transverse magnetization is not exactly vanishing, see Fig.~\ref{fig:transverse}A. To test this hypothesis further, we used numerical simulations in which we slightly rotated the initial product state around the $y$-axis in order to reproduce the experimentally measured transverse magnetization, see Fig.~\ref{fig:transverse}B, finding better agreement with the experiment than with initial states purely along the $x$ direction.

\section{Equilibrium phase diagram\label{sct:numdet}}

In this section, we determine the finite-temperature phase diagram of the model in Eq.~(1) of the main text and discuss how to probe it in the finite-energy setting of the experiment.

\subsection{Finite-temperature phase diagram \label{sec:fintempdiagram}}

To extract the equilibrium properties of the model presented in the main text, we use matrix-product-state (MPS) simulations in the canonical ensemble employing the purification algorithm~\cite{PhysRevB.94.115157,PhysRevB.98.235163} and the $W_{\mathrm{II}}$ matrix-product-operator time evolution method~\cite{PhysRevB.91.165112} implemented in the TeNPy library~\cite{tenpy}. 

\begin{figure}
     \centering
     \includegraphics[width=\columnwidth]{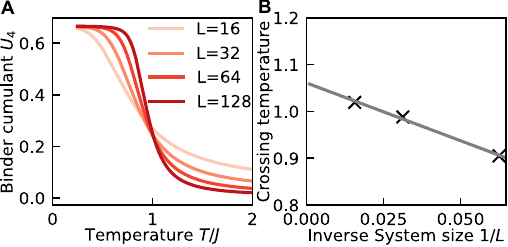}
     \caption{\textbf{Extraction of the critical temperature from the Binder cumulant.} (\textbf{A}) Binder cumulant from purification-MPS for different system sizes. (\textbf{B}) Crossing temperature of consecutive system sizes, where the system size on the x-axis is defined as the smaller of the two system sizes, e.g. the crossing temperature of system sizes $16$ and $32$ is the datapoint at $1/L=0.0625$. Bond dimension $16$, time step $0.01/J$, $\gamma=10.8$, $g/J=0.25$.}
     \label{fig:Binderextract}
 \end{figure}

\begin{figure}
     \centering
     \includegraphics[width=0.9\columnwidth]{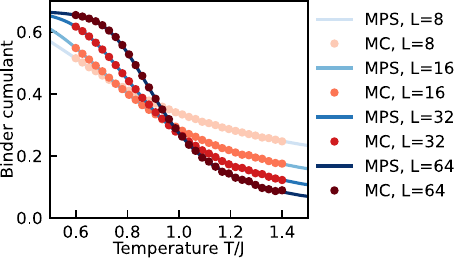}
     \caption{\textbf{Verification of matrix-product-state simulations using classical Monte Carlo.} Monte Carlo simulations employ a Wolff cluster update~\cite{PhysRevLett.62.361} for long-range interactions~\cite{LUIJTEN1995} and $10^6$ Monte Carlo iterations.}
     \label{fig:MC}
 \end{figure}

We employ the standard procedure of using the Binder cumulant
\begin{equation}
    U_4=1-\frac{\braket{(\hat S^x)^4}}{3 \braket{(\hat S^x)^2}^2}
\end{equation}
in order to detect the ferromagnetic phase transition. In this expression, $\braket{\hat S_x^2}\equiv\sum_{ij}\braket{\hat \sigma^x_i\hat\sigma^x_j}/L^2$, $\braket{\hat S_x^4}\equiv\sum_{ijkl}\braket{\hat \sigma^x_i\hat\sigma^x_j\sigma^x_k\sigma^x_l}/L^4$. The Binder cumulant is equal to $2/3$ ($0$) deep in the ferromagnetic (paramagnetic) phase, and the leading finite-size corrections cancel at the phase transition. Hence, calculating the Binder cumulant for consecutive system sizes, extracting their crossing points and extrapolating those to infinite system size leads to a precise estimate for the critical temperature. We show one such procedure in Fig.~\ref{fig:Binderextract} and a comparison of our MPS simulations with a classical Monte Carlo simulation for $g=0$ in Fig.~\ref{fig:MC}. We estimate an error of our extrapolation by comparing $T_c$ obtained from system sizes $16,32,64$ and $16,32,64,128$, finding an error of $0.01-0.03J$.

\begin{figure}
     \centering
     \includegraphics[width=0.8\columnwidth]{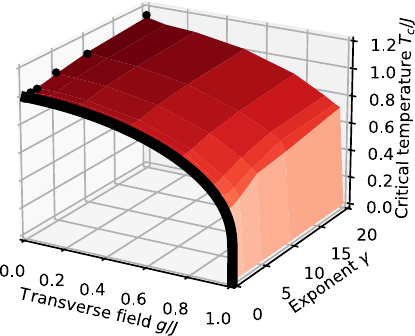}
     \caption{\textbf{Phase diagram extracted from matrix-product-state simulations.} System sizes up to $L=128$ were used in the finite-size extrapolation described in the text. Black dots indicate classical Monte Carlo results for $g=0$. Black solid line is the analytical mean-field result for $\gamma=0$.}
     \label{fig:phasediagram}
 \end{figure}

We show the thus extracted phase diagram in Fig.~\ref{fig:phasediagram}. For $\gamma=0$ (infinite-range interactions), we find excellent agreement with the mean-field solution $T_c/J=(g/J)/\mathrm{arctanh}({g/J})$. Interestingly, we find that $T_c/J$ \emph{increases} as $\gamma/J$ is increased. This indicates that the model shows a phase transition even for large $\gamma/J$. For large $\gamma$, however, increasingly large system sizes are needed to capture the phase transition, which limits the range of our simulations. For $g/J=1$, we find Binder cumulant crossing points which move towards higher temperature as the system size is increased. This indicates that the model is paramagnetic for $g/J>1$ for all $\gamma$ studied here.

\subsection{Scaling}

\begin{figure}
    \centering
    \includegraphics{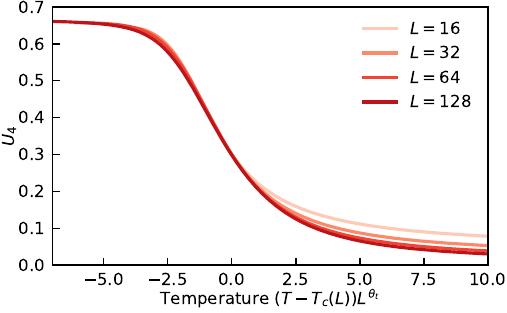}
    \caption{\textbf{Scaling collapse of the Binder cumulant.} Same data as in Fig.~\ref{fig:Binderextract}. $T_c(L)$ is chosen such that all system sizes cross at $T-T_c(L)=0$.}
    \label{fig:collapse}
\end{figure}

We also extract one of the critical exponents from the Binder cumulant by using the method described e.g. in Ref.~\cite{GonzalezLazo2021}: We fit $T_c(L)=T_c(1+a*L^{-\omega-\theta_t})$ to the crossing temperatures and $U_{4,c}(L)=b+cL^{-\omega}$ to the values of the Binder cumulant at the crossing temperatures. From this, we can extract $\theta_t$, which is connected to the critical exponent $\nu$ and find $\theta_t\approx 0.5, 0.52, 0.54, 0.69$ for $g=0.0,0.25,0.5,0.75$. Surprisingly, this indicates a dependence of $\theta_t$ on the transverse field. However, we emphasize that our estimates are rather low-confidence because of the comparatively small system sizes employed here. Moreover, we cannot give error estimates for $y_t$ because we use three data points to fit three parameters. Nevertheless, we show that the thus-found $\theta_t$ leads to a good scaling collapse of the Binder cumulant, see Fig.~\ref{fig:collapse} and compare to Fig.~\ref{fig:Binderextract}A for the uncollapsed data. The exponent $\theta_t=0.5$ corresponds to the mean-field exponent which is exact in the all-to-all connected model $\gamma=0$. We also extract a slightly different $T_c$ from this procedure, but generally find agreement with the procedure shown in Fig.~\ref{fig:Binderextract} within about 3\%.

\subsection{Energy phase diagram from canonical simulations\label{sct:Microcanphasediagram}}

\begin{figure}
    \centering
    \includegraphics{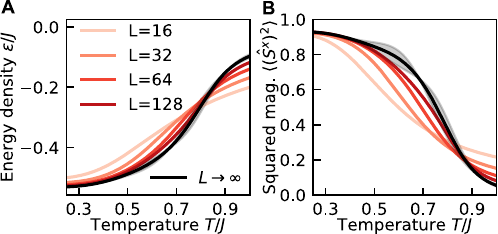}
    \caption{\textbf{MPS results for temperature-dependent observables.} (\textbf{A}) Energy density and (\textbf{B}) squared magnetization. The infinite-system-size extrapolation is done using the system sizes shown, using a quadratic fit to the finite size results. $g/J=0.25$. Gray shading is an estimate for the residual finite size error obtained by comparing the extrapolation result using system sizes $16,32,64$ and $16,32,64,128$.}
    \label{fig:energy_temp}
\end{figure}

We extract the finite-energy/microcanonical properties of the model by using the equivalence of ensembles in the infinite-system-size limit. To do so, we extrapolate the energy as a function of temperature $E(T)$ to infinite system size, see Fig.~\ref{fig:energy_temp}A. Doing the same for observables, see Fig.~\ref{fig:energy_temp}B, enables us to convert all observables to be as a function of energy instead of temperature by similarly extrapolating the value of the observables to infinite system size. In particular, we obtain the critical energy line shown in Fig. 1C of the main text by using the extrapolated numerically obtained $E(T)$ to obtain the critical energy $E_c/J$ from the critical temperature $T_c/J$, which was obtained by the procedure discussed in Sec.~\ \ref{sec:fintempdiagram}.

\begin{figure}
    \centering
    \includegraphics{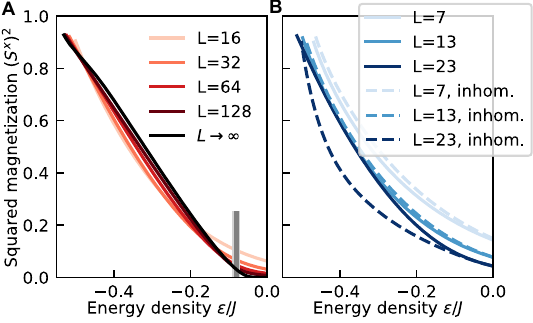}
    \caption{\textbf{Finite-size behaviour of energy-dependent squared magnetization.} Solid lines are obtained from MPS simulations in the canonical ensemble using the target interactions defined in Eq.~(1) of the main text. Transverse field $g/J=0.25$, $\gamma=10.8$, bond dimension $\chi=16$. (\textbf{A}) System sizes chosen as powers of two. The vertical grey line indicates the critical energy $E_c/J\approx -0.08$ obtained as described in section~\ref{sct:Microcanphasediagram}. The thickness of the light grey line indicates the uncertainty of the critical energy due to residual finite-size effects, defined as the difference in critical energy obtained from the infinite-system-size extrapolation using system sizes $16,32,64$ and $16,32,64,128$. Dark grey shading is an independent estimate of the uncertainty resulting from the uncertainty in the critical temperature ($\pm 0.01J$). (\textbf{B}) Experimentally probed system sizes using homogeneous interactions (full lines) using the inhomogeneous experimentally realized interactions, shown in Fig.~\ref{fig:expinteractions}.}
    \label{fig:squmag_E}
\end{figure}

\begin{figure}
 \centering    \includegraphics[width=0.8\columnwidth]{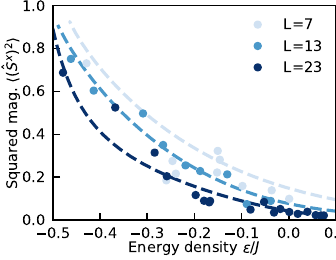}
\caption{\textbf{Experimentally measured squared magnetization for different system sizes.} Experimentally measured late-time squared magnetization (dots) compared to the numerical results in the canonical ensemble for the experimental $J_{ij}$ (dashed lines, same as dashed lines in Fig.~\ref{fig:squmag_E}B). $g/J=0.24,0.21,0.18$ for $L=7,13,23$, respectively.}
     \label{fig:5}
 \end{figure}
\begin{figure*}
    \centering
    \includegraphics{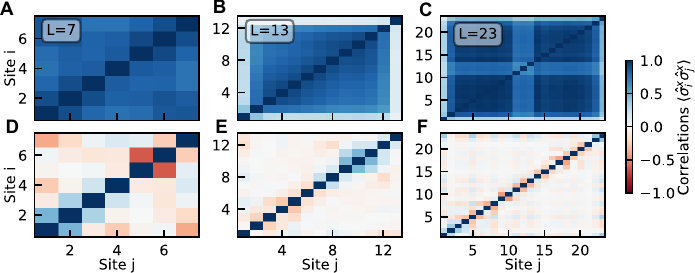}
    \caption{\textbf{Correlation functions from numerics.} Same parameters as Fig.~4 in the main text.}
    \label{fig:supp_4}
\end{figure*}
\begin{figure*}
     \centering
 \includegraphics{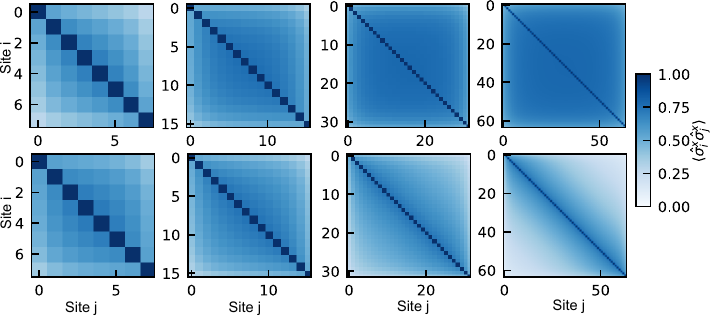}
     \caption{\textbf{Correlation function.} Matrix-product-state simulations. $g/J=0.5$, temperature $T/J=0.25$, bond dimension $16$, time step $0.01/J$. System size $8,16,32,64$ increasing from left to right. In the top row [bottom row] we use the interactions $J_{ij}$ in the main text [$J_{ij}^\mathrm{unnormalized}$ defined in Eq.~\eqref{eq:Jij_unnorm}]. We find that the correlations decay faster as system size is increased when the interactions are not normalized, indicating the absence of long-range order.}
     \label{fig:corrfct}
 \end{figure*}
\begin{figure}
     \centering
     \includegraphics[width=\columnwidth]{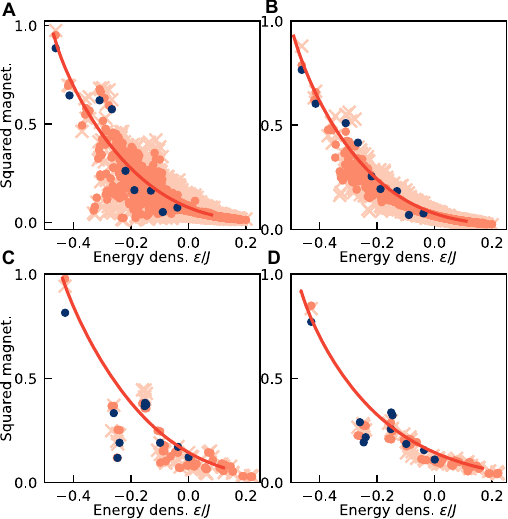}
     \caption{\textbf{Thermalization of all x-product initial states.} Obtained from exact diagonalization evolved to the experimental latest time (crosses) and infinite time (orange dots). Blue dots are the latest time points of the time averaged experimental data. (A) L=13, $g/J=0.1$, (B) L=13, $g/J=0.21$, (C) L=7, $g/J=0.12$, (D) L=7, $g/J=0.24$ }
     \label{fig:diag}
 \end{figure}

\subsection{Finite-size effects in energy-dependent observables}

The finite-size dependence of temperature-dependent observables in the vicinity of the critical temperature of a second-order phase transition can be captured by considering the growth of the correlation length as the system size increases, leading to the theory of finite-size scaling. However, much less is known about finite-size scaling in the microcanonical ensemble~\cite{Desai1988} and even less in the diagonal ensemble, which is the one experimentally realized in this work. To get a rough idea for how strong finite-size effects are on energy-dependent observables, we display the squared magnetization for different system sizes as a function of energy, evaluated with MPS simulations in the \emph{canonical} ensemble in Fig.~\ref{fig:squmag_E}A. We find that the finite-size corrections for $L\gtrsim 32$ are almost invisible for this observable and much weaker than, for instance, the finite-size corrections in the Binder cumulant as a function of temperature shown in Fig.~\ref{fig:Binderextract}. This is in tune with the observation that the measured squared magnetization for $L=13$ shown in Fig.~3 of the main text is close to the infinite-system-size extrapolation shown in Fig.~1C. We found that finite-size effects become stronger as $g/J$ is increased. Moreover, we find in Fig.~\ref{fig:squmag_E} that the squared magnetization extrapolated to infinite system size does not exactly vanish at the critical energy extracted from evaluating the infinite-system-size extrapolated energy at the critical temperature. We attribute this to the fact that the infinite-system-size extrapolation has not fully converged yet for both the critical energy (c.f. the grey error bar lines in Fig.~\ref{fig:squmag_E}) and also the squared magnetization in the vicinity of the critical energy (c.f. the uncertainty for both the energy density and the squared magnetization in Fig.~\ref{fig:energy_temp}. We note that this is to be expected for finite-size extrapolation of the order parameter in an Ising-like transition as in a finite-size system the order parameter ``rounds off'', leading to a small, but non-vanishing magnetization even for $\epsilon/J>\epsilon_c/J$ in finite systems.

To test how experimental imperfections in the interactions due to inhomogeneities affect this result, we compare  the canonical result using the calculated experimentally realized interactions (c.f. Fig~\ref{fig:expinteractions}) with the corresponding result using the ideal, homogeneous interactions. We find that the inhomogeneities in the interactions lead to deviations from the homogeneous result, which hampers a systematic finite-size extrapolation. Nevertheless, as shown in Fig.~\ref{fig:5},  the experimental data, which is data obtained from time evolution, shows little finite-size effects as expected from our numerical calculations for homogeneous interactions. Moreover, we note that in principle the interactions could be made homogeneous by further engineering the addressing beams~\cite{Korenblit2012}.

\subsection{Correlation functions}

In Fig.~\ref{fig:supp_4} we show numerically calculated correlation functions at high and low energy with the same parameters as in Fig.~4 of the main text. We evolved to the same time as in the experiment and used the calculated experimentally realized interactions shown in Fig.~\ref{fig:expinteractions}. We find in general good agreement. In particular, the remnant of ferromagnetic correlations in the top right corner of Fig.~\ref{fig:supp_4}E and F is also visible in the experimental result, showcasing that the measured patterns are indeed physical. Moreover, the slight reduction in correlations in the centre of Fig.~\ref{fig:supp_4}C is also visible in the experiment, albeit is less pronounced.

In order to show the crucial role of normalizing the exponent of the exponential decay of the interactions by $1/L$, we show in Fig.~\ref{fig:corrfct} the correlation function $\braket{\hat \sigma^x_i \sigma^x_j}$ for the definition of the Hamiltonian in the main text (i.e. with perfectly homogeneous interactions) as well as without the normalization, i.e.
\begin{equation}
    J_{ij}^\mathrm{unnormalized}=\exp\left(-\frac{\gamma}{13}|i-j|\right),
    \label{eq:Jij_unnorm}
\end{equation}
with $\gamma=10.8$, where we chose the denominator in order to match the definitions for the relatively small system size $L=13$. We find that, while the correlations become longer ranged as the system size increases for the definition in the main text, the correlations decay to zero quickly as the system size increases when using $J_{ij}^\mathrm{unnormalized}$, showing the absence of a long-range-ordered phase at this temperature. This is in agreement with the expectation that purely exponentially decaying interactions are in the universality class of the model with nearest-neighbour interactions.

\begin{figure*}
     \centering
     \includegraphics[width=4.75in]{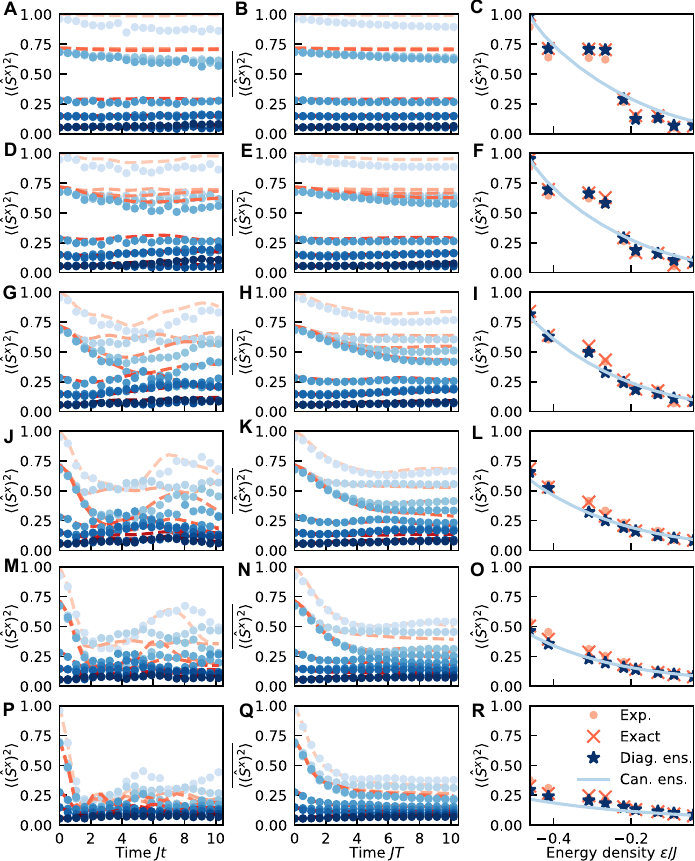}
     \caption{\textbf{Verification of equilibration.} Same plot as Fig.~2 of main text for all transverse fields shown in Fig.~3 of the main text. The transverse field is fixed for each row and increases from top to bottom: $g/J=0.04, 0.10, 0.21, 0.31, 0.41,
       0.62$. $g/J=0.31J$ corresponds to Fig.~2 of the main text, which we repeat here for completeness. System size $L=13$. First column: time-evolved squared magnetization in the experiment (dots) and numerical simulations (dashed lines). Second column: time-average (up to time $T$) of the time-evolved squared magnetization using the data from the first row (dots) and the corresponding numerical data (dashed lines), evaluated according to Eq.~2 of the main text. Third column: comparison of the latest-time experimental data points from the second column (dots), numerical data evolved until the experimental time (crosses) and to infinite time, i.e. the diagonal ensemble (stars, evaluated according to the RHS in Eq.~\eqref{eq:timeaver}). The expectation from the canonical ensemble is shown as a solid line. The numerics use the experimentally realized interactions. $L=13$, $g=0.31J$. Error bars for the experimental data are quantum projection noise and are smaller than the point size.}
     \label{fig:S_timeevol}
 \end{figure*}

 \begin{figure*}
     \centering
 \includegraphics{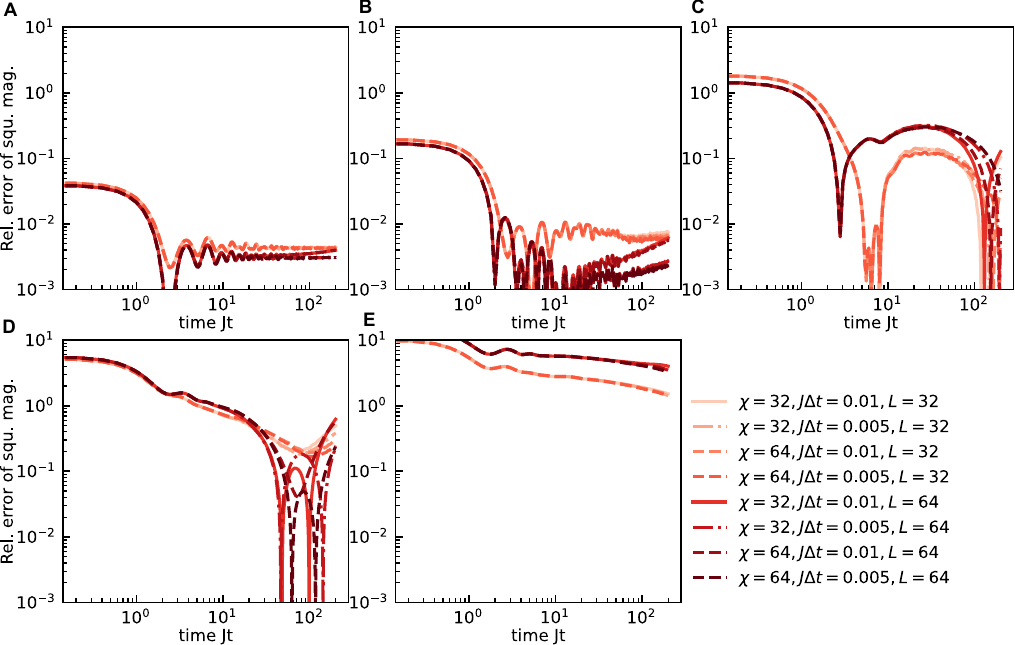}
     \caption{\textbf{Test of thermalization for large systems.} Matrix-product-state simulations for the totally polarized initial state using bond dimension $\chi$ and imaginary time evolution step $\Delta t$. $\gamma=10.8$. $g/J=0.125,0.25,0.5,0.625,0.75$ for (\textbf{A}), (\textbf{B}), (\textbf{C}), (\textbf{D}), (\textbf{E}), respectively.}
     \label{fig:Testtherm}
 \end{figure*}

\section{Time-evolution method to evaluating equilibrium observables}
In this section, we introduce our method in greater detail and check several underlying assumptions for our specific case.

\subsection{Overview of the method}

Our method to experimentally prepare thermal states is using a measurement of time-averaged observables $\hat O$

\begin{align}
\overline{\braket{\psi| \hat O(t) |\psi} }&\equiv \int^T_0 dt \braket{\psi| \hat O(t) |\psi} \notag\\&\stackrel{T\rightarrow \infty} {\rightarrow}\sum_{n} |\braket{n|\psi}|^2 \braket{n|\hat O|n}\label{eq:timeaver}
\end{align}
with respect to an initial state $\ket{\psi}$ of energy $E=\braket{\psi|\hat H|\psi}$. In the second step, we inserted the energy eigenbasis $\hat H \ket{n}=E_n\ket{n}$ and assumed that the Hamiltonian has a non-extensive number of degenerate states. 
The eigenstate thermalization hypothesis~\cite{Srednicki1996} states that the expectation values of a local observable $\hat O$ with respect to eigenstates $\ket{n}$ of the Hamiltonian with eigenenergy $E_n$ are given by
\begin{equation}
    \braket{n|\hat O|n}=\mathcal{O}(E_n), 
\end{equation}
where $\mathcal{O}(E_n)$ is the value of the observable in the microcanonical ensemble at energy $E_n$, which is a smooth function of energy. If the probability distribution $|\braket{n|\psi}|^2$ has a vanishing energy density variance $(\braket{\psi|\hat H^2|\psi}-\braket{\psi|\hat H|\psi}^2)/L^2 \stackrel{L\rightarrow \infty} {\rightarrow} 0$ in the thermodynamic limit, then we can directly follow from ETH that $\sum_{n} |\braket{n|\psi}|^2 \braket{n|\hat O|n}=\mathcal{O}(E) \sum_{n} |\braket{n|\psi}|^2=\mathcal{O}(E)$ and therefore

\begin{equation}
    \overline{\braket{\psi| \hat O(t) |\psi} } \rightarrow \mathcal{O}\left(\braket{\psi|\hat H|\psi}\right).
\end{equation}

 This motivates the following protocol to evaluate expectation values of observables in the microcanonical ensemble, which we follow in the main text:\\
(1) Prepare initial state $\ket{\psi}$, e.g. a product state, with energy $E_\psi=\braket{\psi|\hat H|\psi}$.\\
(2) Evolve under Hamiltonian $\hat H$ until time $t$.\\
(3) Measure $\hat O$ at various times $t\leq T$.\\
(4) Evaluate the time average in Eq.~\eqref{eq:timeaver}. \\
(5) Converge with respect to $T$.

\subsection{Finite-size effects of the diagonal ensemble}

In order for the diagonal-ensemble expectation values (i.e. the late-time expectation values of observables) to approach the microcanonical ensemble, the expectation values of the observables with respect to the eigenstates $\braket{n|\hat O|n}$ should form a smooth function of the energy $E_n$ over the width of the diagonal ensemble $|\braket{n|\psi}|^2$. In a finite system, both of these conditions can be and often are violated. This means that the long-time expectation value of an observable evolved from two initial states that are close in energy can be very different. We show this observation for our system in Fig.~\ref{fig:diag}, where we show the diagonal ensemble expectation values along with the time-averaged expectation value of the observable evolved to the latest time available in the experiment for all $2^L$ x-product states, as well as the canonical ensemble and the experimental data. We find that, for small transverse fields, the scatter around the canonical ensemble is larger, which we attribute to the interplay between the weak breaking of integrability at $g=0$ and finite-size effects. We also find that the time we evolved to in the experiment is largely long enough to reproduce the overall behaviour as a function of energy, with some finite-time differences visible for $g/J=0.21$ and $L=13$.

We also note that the states we chose in the experiment by the procedure described in section~\ref{sct:stateprep} are roughly representative of all states, i.e. their scatter around the canonical ensemble is similar to the scatter of all product states.

\subsection{Verification of thermalization}
In Fig.~\ref{fig:S_timeevol}, we show the analogue of Fig.~2 in the main text across all transverse fields shown in Fig.~3 of the main text, with the same initial states for all fields, tabulated in table~\ref{tab:states13}. We find a similarly good match for all fields, with the agreement becoming better for larger fields. For the numerical calculations, we employed the interactions shown in Fig.~\ref{fig:expinteractions}B. 

To test thermalization more rigorously, we perform matrix-product-state simulations for larger systems, starting from the totally polarized initial state $\ket{\uparrow \cdots \uparrow}$, defined along the x-axis (i.e., the lowest-energy product initial state for $g=0$). We show the relative deviation from the thermal expectation value (obtained from purification-MPS simulations, as introduced in section~\ref{sct:numdet}, at the same system size) in Fig.~\ref{fig:Testtherm}. We find that, for small fields, the relative error reaches a plateau as a function of time, with the value of the plateau decreasing for increasing system size, which is an indication that, in the infinite-time and infinite-system-size limit, the long-time value and the thermal value approach each other. In general, the relative error increases with transverse field. At around $g/J\approx 0.625$, we find an approximately power-law decay of the relative error instead of a plateau, before erratic oscillations set in that depend heavily on the time step and bond dimension. For even larger fields, a plateau is again reached which however now tends to larger error as the system size is increased. This is to be expected as, for large transverse fields, this model shows prethermalization: instead of a thermalization to the transverse-field Ising model, the system thermalizes to the XY model due to the quasi-conservation of the transverse field. Indeed, at large transverse field, the long-time value of the squared magnetization approaches $1/2$ as expected from the XY model instead of $0$ as expected from the transverse field Ising model. Moreover, at intermediately large field, this deviation from thermal behaviour has been associated to a dynamical quantum phase transition~\cite{Zhang20172}, but the exact relation to prethermalization is unresolved. In any case, these effects constrain the applicability range of our method to small values of the transverse field.

\bibliography{lit.bib}